\RequirePackage{lineno}
\documentclass[%
 reprint,
 prd,
 amsmath,amssymb,
 aps,
 lengthcheck,
floatfix
]{revtex4-2}
\usepackage[overload]{textcase}
\usepackage{graphicx}
\usepackage{dcolumn}
\usepackage{bm}
\usepackage[colorlinks=true,linktocpage=true,linkcolor=blue,citecolor=blue,allcolors=
blue] {hyperref }
\usepackage[utf8]{inputenc}
\usepackage{multirow}
\usepackage{soul}
\usepackage{float}
\usepackage{graphicx}
\usepackage{times}
\usepackage[normalem]{ulem}
\usepackage{color}
\usepackage{cellspace}
\usepackage{lipsum}
\newcommand{\bef}{\begin{figure}}
\newcommand{\eef}{\end{figure}}
\newcommand{\bc}{\begin{center}}
\newcommand{\ec}{\end{center}}

\newcommand{\antinue}{\ensuremath{\overline{\nu}_{e}}}
\newcommand{\comment}[1]{}
\renewcommand\sout{\bgroup \color{red} \ULdepth=-.5ex \ULset}

\begin{document}
\title{Sensitivity study of a sapphire detector using Coherent Elastic Neutrino-Nucleus Scattering process}

\author{S. P. Behera$^{1}$$^{,2}$}
\email{shiba@barc.gov.in}
\affiliation{$^{1}$Nuclear Physics Division, Bhabha Atomic Research Centre,\\ Mumbai - 400085, India}
\affiliation{$^{2}$Homi Bhabha National Institute, Anushakti Nagar, Mumbai - 400094, India}
\begin{abstract}
The Indian Coherent Neutrino-nucleus Scattering Experiment(ICNSE)  
has been proposed at Bhabha Atomic Research Centre in India to measure
 the coherent elastic neutrino-nucleus scattering process using electron
  antineutrinos produced from reactors. Phenomenological studies are 
  performed to find out the sensitivity of a sapphire detector for various
   fundamental physics parameters at an exposure of one year. Reactors 
   of different core compositions, sizes, and thermal powers have been 
   considered as sources of electron antineutrinos. The potential of the
    ICNSE to measure the weak mixing angle at a low energy regime has been
     extracted. Furthermore, the detector's capability has been investigated for examining 
the electromagnetic properties of neutrinos, including their magnetic moment. 
Additionally, an exploration has been conducted on the detector's sensitivity 
in restricting new interactions between neutrinos and electrons or nuclei, 
thereby constraining the parameter space related to light mediators. 
It is found that the ICNSE detector can put a stronger constraints on the scalar and vector mediators masses.
\end{abstract}
%%%%%%%%%%%%%%%%%%%
\maketitle
%%%%%%%%%%%
\section{Introduction}
\label{sec:intro}
The concept of coherent elastic neutrino-nucleus scattering (CE$\nu$NS) was initially
 introduced by Freedman~\cite{Freedman:1973yd} within the framework of the standard model (SM) of particle physics.
The CE$\nu$NS process involves low-energy neutrinos scattering off the entire atomic nucleus
 via neutral-current interactions within the SM of weak interactions.
For low momentum transfer, the CE$\nu$NS cross section is approximately proportional
 to number of neutrons present in the target nuclei.
In the CE$\nu$NS process, the nuclei that are scattered carry energy on the order of keV.
Measurement of such low energy recoil nuclei is extremely difficult.
In contrast, this interaction channel provides a significantly higher interaction rate per target atom, 
ranging from 3 to 4 orders of magnitude, when compared to other detection methods like inverse
 beta decay and neutrino electron scattering~\cite{CONUS:2021dwh}. As a result, it enables a 
significantly smaller target size than the traditional neutrino detectors. Conversely, an experiment
 with a large target mass can obtain high neutrino statistics, allowing for precise measurements
 of various neutrino parameters.

The COHERENT group has recently conducted the first measurements on the CE$\nu$NS process~\cite{COHERENT:2017ipa}
with accelerator neutrinos at the Spallation Neutron Source.
They observed the process at a 6.7$\sigma$ confidence level using a low-background
CsI[Na] scintillator.
In the following
years, the group reported a second measurement in an
LAr detector~\cite{COHERENT:2020iec}, and another data set from the CsI
measurement has been released~\cite{COHERENT:2021xmm}. The measured cross section is consistent
with the standard model prediction.
They also have the first-ever detection of the CE$\nu$NS process on germanium
 nuclei with a significance of 3.9$\sigma$~\cite{COHERENT:2024axu}.
The measurement of the CE$\nu$NS cross-section provides a pathway to explore a variety 
of physics phenomena across different research fields including particle physics, 
nuclear physics,  astrophysics, and cosmology. Studying the CE$\nu$NS 
process can shed light on fundamental aspects of physics beyond the SM, such as the non-standard 
interactions~\cite{Papoulias:2017qdn}, the neutrino magnetic moment~\cite{Kosmas:2015sqa},  
the weak mixing angle~\cite{Canas:2017umu,Papoulias:2017qdn}, and the nuclear neutron 
density distributions. Furthermore, the possible existence of sterile neutrinos
 could be confirmed or disproved by observing neutrinos through the flavor- 
 blind (CE$\nu$NS) process. This process also allows for detailed investigations 
 into the interiors of dense objects and stellar evolution~\cite{Biassoni:2011xuo,Brdar:2018zds}.
The CE$\nu$NS process not only offers to explore the Beyond Standard 
Model (BSM) physics but also can be applied for monitoring nuclear reactors.
In India, an experimental setup the Indian Coherent Neutrino-nucleus Scattering 
Experiment(ICNSE),  has been proposed to measure the CE$\nu$NS 
cross section using electron antineutrinos produced from the reactor and 
address various fundamental physics aspects.

The current study examines the detection capabilities of the ICNSE detector
 for measuring the weak mixing angle, neutrino magnetic moment, and masses 
 of various mediators. The weak mixing angle has been measured at higher 
 energy in GeV scale~\cite{CHARM-II:1994dzw}, while the measurement at low 
 energy requires improved accuracy of this parameter~\cite{Erler:2004in,Canas:2016vxp}.
The cross section for the CE$\nu$NS depends on the weak charge, aiding in the 
study of the weak mixing  angle at extremely low momentum transfer.  With 
the established fact that neutrinos possess non-zero mass from the neutrino 
oscillation experiment, they are expected in the extension of
 the SM to exhibit electromagnetic properties such as the neutrino magnetic
  moment and the neutrino charge radius.
The nuclear recoil energy spectrum can be distorted due to neutrino 
interaction with nuclei in the presence of a magnetic moment. Further 
 neutrinos coupling to protons and neutrons may be potentially influenced 
 due to the non-standard interaction of neutrinos. Then it introduces 
 additional new parameters that can provide insight into the relative 
 magnitude of these interactions compared to the neutral-current weak 
 interaction as described in the the SM~\cite{Giunti:2019xpr}.  These 
 couplings can be expressed  as a function of the  mediator mass and the momentum transfer.
In case of a momentum transfer greater than the mediator mass, the 
couplings exhibit transfer momentum dependence, which leads to the 
possibility of new physics phenomena, such as the introduction of a 
new gauge symmetry featuring an additional scalar or vector mediator.
 A scalar or vector particle can participate in CE$\nu$NS process and 
 potentially alter the nuclear recoil spectrum in a unique way. 
 The characteristic distortion of the spectrum shape for light scalars
  with masses around the neutrino energy allows us to reconstruct  the 
  scalar mass. The COHERENT group has put a limit on its mass and coupling with SM particles~\cite{COHERENT:2017ipa}.
 
The article is organized as follows. In the following section, a detailed 
description of the proposed ICNSE setup is presented. The production 
mechanism of neutrinos inside the reactor is elaborated in Sec.~\ref{sec:reactor}. The 
CE$\nu$NS process and the principle of measurement  are discussed in 
Sec.~\ref{sec:cens}. The procedure for estimating the 
expected number of events in the detector is described in Sec.~\ref{sec:expect}. 
  The sensitivity of the proposed experiment and statistical method  
  on $\chi^{2}$ estimation considered in this study are discussed in 
  Sec.~\ref{sec:simul}. The sensitivity of the detector 
  to various physics parameters at an exposure of one year is elaborated 
in Sec.~\ref{sec:results}. In Sec.~\ref{sec:summary},  observations 
obtained from this study are summarized, and the implication of this work are discussed.
%%%%%%%%%%%%%%%%%%%%%% Fig.1 %%%%%%%%%%%%%%%%%%%%%%%%%%%%%%%%%
%%%%%%%%%%%%%%%%%%%%%%%%%%%
\begin{figure}%[t]
\advance\leftskip -50cm
\centering
\includegraphics[trim={3cm 0 10cm 0cm}, width=0.4\linewidth]{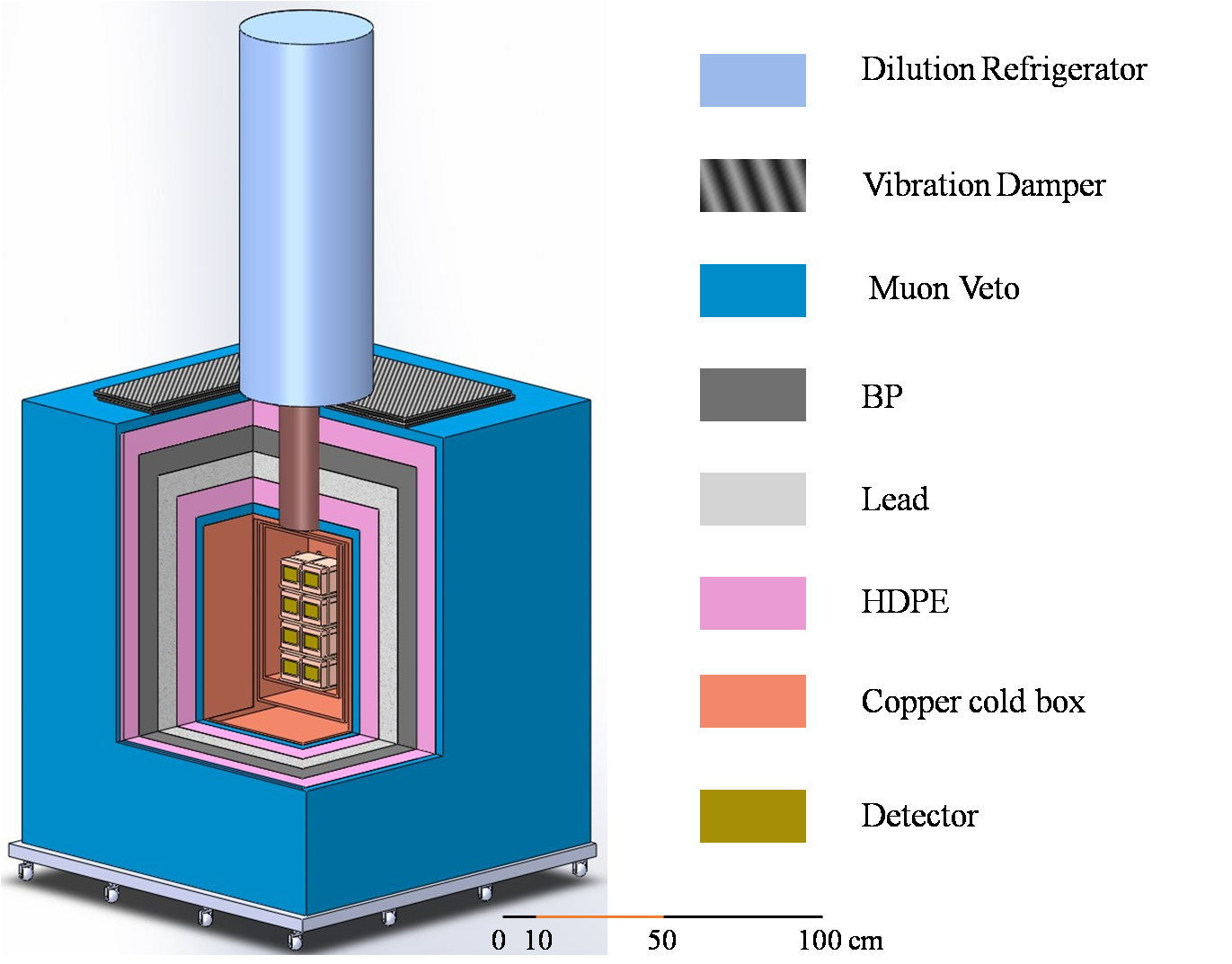}
\caption{ \label{fig:DetSetup} Schematic representation of the ICNSE experimental setup
for CE$\nu$NS process measurement. BP: borated polyethylene, thickness: 10 cm; HDPE: high-density polyethylene,thickness: 10 cm.
The figure has been taken from Ref.~\cite{Behera:2023llq}.} 
\end{figure}
%%%%%%%%%%%%%%%%%%%%%%%%%%%%%%%%%%
\begin{table*}[t]
 \begin{center}
\caption{\label{tab:reactortype}{Various types of reactors used as $\antinue$ sources }}
\begin{tabular}{ cccc}
    Reactors name & Thermal power(MW$_{th}$) & Fuel type & Core sizes, R: radius, H: Height\\
\hline
   Apsara-U & 3.0 & U$_{3}$Si$_2$-Al (17$\%$  enriched $^{235}$U) & R = 0.32 m, H =0.64 m \\
    Dhruva & 100.0 & Natural uranium (0.7$\%$  $^{235}$U) & R = 1.5 m, H =  3.03 m \\
 PFBR & 1250.0 & MOX(PuO$_{2}$-UO$_{2}$)  & R = 0.95 m, H =  1.0 m \\ 
 VVER & 3000.0 &  UO$_{2}$ (3.92 $\%$ enriched $^{235}$U) & R = 1.58 m,  H = 3.53 m \\
   \hline
\end{tabular}
\end{center}
\end{table*}
%%%%%%%%%%%%%%%%%%%%%%%%%%%%%%%%%%%%%%%%%%%%%%%%%%%%%
%%%%%%%%%%%%%%%
\section{Proposed ICNSE Detector Setup for the CE$\nu$NS Process Measurement }
\label{detSetup}
%%%%%%%%%%%%%%
%%%%%%%%%%%%%%%%%%%%%%%%%%%%%
As mentioned earlier that, experiments based on CE$\nu$NS process require detectors with very low threshold 
that are essential for the measurement of very
 low nuclear recoil energies ($\sim$few keV). Semiconductor detectors based
on germanium or silicon are good candidates to achieve the very low threshold required. 
However, the energy thresholds of such detectors are $\mathcal{O}$(keV) which is
also very high to observe  the CE$\nu$NS process using reactor antineutrinos. Recoil energies can be measured using cryogenic
  detectors, which can achieve excellent energy resolutions along with low thresholds. 
 Cryogenic detectors can be made from a lot of different materials,
such as semiconductors like germanium or insulators like calcium tungstate 
and sapphire($Al_{2}O_{3}$).
However, sapphire is a very good candidate to observe the CE$\nu$NS
due to the lower atomic mass of $Al$ and $O$,  making it sensitive to lower nuclear recoil energies.
Sapphire
has good phononic properties and it has already been shown that low energy
thresholds $\mathcal{O}$(0.1keV) are possible with these crystals.
A baseline recoil energy resolution of 18 eV that corresponds to a recoil energy threshold of 54 eV
has been achieved using a scale of 100 g newly developed sapphire detector~\cite{Verma:2022tkq}.
According to a sensitivity study carried out in Ref.~\cite{Strauss:2017cuu},
 the CE$\nu$NS process can be measured at a 5$\sigma$ level using cryogenic 
 detectors with recoil energy thresholds of 20 eV.
 Thus, for the present work, only sapphire has been
used as a target material. The conceptual design of the detector setup has been considered
from Ref.~\cite{Behera:2023llq} which is shown in Fig~\ref{fig:DetSetup}. 

 Additionally, the assessment of low nuclear recoil energy is influenced 
 by background noise originating from both the reactor core and environmental 
 factors. These background sources encompass gamma rays and neutrons emitted 
 by the reactor, as well as muons and muon-induced neutrons resulting from 
 cosmic rays and surrounding gamma radiation from natural radioactivity. The energy-dependent backgrounds 
 attributed to neutrons and gamma rays are evaluated in Ref.~\cite{MINER:2016igy}.
 Also, several rare event search experiments observe sharply rising backgrounds at sub-keV energies 
 close to their respective thresholds that are larger than expected from known backgrounds~\cite{Fuss:2022fxe}.
   Therefore,  it is essential to suppress natural and reactor related backgrounds comprising mostly of gamma-rays, 
neutrons, and muons. This is achieved by using a multilayer structure shielding materials as 
shown in Fig.~\ref{fig:DetSetup}. The outermost layer 
of this shielding is a 4 cm thick plastic scintillator (PS) for vetoing external radiation. This is followed 
by  10 cm thick layers of high-density polyethylene (HDPE) and borated polyethylene (BP) sheets
 containing 15$\%$ boron for thermalizing fast neutrons and subsequently attenuating them. 
 This is followed by  10 cm layer of lead shielding to attenuate gamma radiation. An additional 
 layer of 10 cm thick HDPE followed by a 4 cm thick PS are placed inside the 
 lead layers to thermalize and tag any fast neutrons produced in the lead layer via ($\gamma$,n) 
 reactions or muon-induced neutrons. The detectors are housed in an 
 oxygen-free high thermal conductivity (OFHC) copper
 cold box placed at the center of this shielding. The copper box also shields the detector 
 from charged radiation. The thickness of lead and BP sheets was decided considering 
 an earlier study ~\cite{Mulmule:2018efw} carried out for the Indian Scintillator 
 Matrix for Reactor Anti-Neutrino, an inverse beta decay reaction based reactor antineutrino experiment.
 To maintain the required cryogenic temperatures(about 15 mK  for $\alpha$-phase of W ~\cite{Jastram:2014pga}) 
 for the detector operation, a dilution refrigerator is placed on top
 of the shielding. A cold finger will establish the thermal link for cooling the cold box from the dilution refrigerator. 
  It is also important to take the precaution of ensuring proper thermal link for stable operation while damping out 
  vibrations from reaching the detector setup. The signal extraction method from the sapphire detector is described below. 
      
  In the CE$\nu$NS process process, the recoiling nucleus induces athermal excitation in the detector crystal,
   with approximately 92$\%$ of the energy being converted into generating athermal phonons and 
   remaining energy goes for producing scintillating photon signals. Consequently, the recoil 
   energy of nuclei can be determined by monitoring the signals stemming from phonons. 
   Additionally, the phonon signals are not contingent on the nature of the interaction $i.e$ signals 
   due to either electron recoil or nuclear recoil as the detector is a calorimetric device~\cite{Heikinheimo:2021syx,Strauss:2017cuu}.
     From the ICNSE experimental setup, phonons will be identified through a dual-stage 
   procedure employing quasiparticle-trap-assisted electrothermal-feedback transition-edge-sensors made up of Al fins 
coupled to tungsten Transition Edge Sensors(W-TES), fabricated on the surface of the detector.
 The measurement process involves the collection of phonons by the aluminum fins, 
 leading to the disruption of cooper pairs and the generation of quasi-particles.
  Subsequently, these quasi-particles disperse throughout the aluminum fins until
   they are captured in the overlap region, where they enter into tungsten, 
   inducing a rise in temperature. By maintaining tungsten within its superconducting 
   transition range, this temperature fluctuation results in a significant alteration
    in resistance, which is quantified by applying a current through the W-TES under
     voltage biased mode~\cite{Verma:2022tkq}.   
%%%%%%%%%%%%%%%%%%%%%%%%%%%%%%%%% 
%%%%%%%%%%%%%%%%%%%%%%%%%%%%%%%%%%%%%%%%%%%%%%%%%%%%%%%%%%%%%%%%%%%%%%%%%%%%%%%%
\section{Production of Electron Anti-neutrinos from the Reactor}
\label{sec:reactor} %                            %
%%%%%%%%%%%%%%%%%%%%%%%%%%%%%%%%%%%%%%%%%%%%%%%%%%%%%%%%%%%%%%%%%%%%%%%%%%%%%%%%
Man-made nuclear reactor are intense sources of pure \antinue s.
 On average, 10$^{20}$ \antinue s are produced from a 1 GW$_{th}$ thermal
    power reactor. In a reactor,  \antinue s are produced mainly by two
     processes. One of them is the beta decay of neutron-rich fission fragments of 
     mainly four isotopes, such as $^{235}$U, $^{238}$U, $^{239}$Pu, and $^{241}$Pu.  
     Each isotope has a different fission rate which leads to different  \antinue s~ yield
      and spectrum. \antinue s have a maximum energy of about 10 MeV, which is 
     produced from the beta decay of fission fragments. Another important 
     one is from the neutron capture process by the $^{238}$U that leads to the production of two  \antinue s
having energy $<$1.3 MeV~\cite{FernandezMoroni:2014qlq}. This process 
accounts for approximately 16$\%$ of the overall \antinue flux.
These low energy  neutrinos are usually neglected, as most of the experiments detect \antinue
through the inverse beta decay process. It can be noted here that the relative contribution of each isotope 
to the  total \antinue s flux depends on the fuel composition of the reactors 
and their burning cycle.  There is also a small variation in flux that occurs due to 
fuel burn-up. The present study has been carried out considering various types of
 reactors with different core sizes, compositions, and 
 thermal power. Details on reactor thermal power, compositions, and core sizes are given
  in Table~\ref{tab:reactortype}. At present, it is planned to put the detector at 4 m from the reactor core in 
the Apsara-U research reactor facility at Bhabha Atomic Research Centre 
(BARC), India~\cite{SINGH2013141}.  The main advantage of the Apsara-U reactor
 is that it has a movable and compact core. The utilization of a mobile core 
 offers the benefit of mitigating systematic uncertainties associated with the 
 reactor and the detector through the implementation of position-specific measurements.
 In the future, the same detector setup can be placed at other reactor facilities 
 such as Dhruva, BARC~\cite{AGARWAL2006747},  proto-type fast breeder reactor 
 (PFBR), IGCAR, Kalpakkam~\cite{CHETAL2006852}, and VVER, Kudankulam in India~\cite{AGRAWAL2006812}. 
%%%%%%%%%%%%%%%%%%%%%%%%%%%%%%%%%%%%%%%%%%%%%%%%%%%%%%%%%
\section{Measurement of the coherent neutrino nucleus scattering process}\label{sec:cens}
%%%%%%%%%%%%%%%%%%%%%%%%%%%%%%%%%%%%%%%%%%%%%%%%%%%%%%%%%
The measurement of the CE$\nu$NS process provides a probe to study the BSM physics. 
The differential CE$\nu$NS scattering cross-section is given by
%%%%%%%%%%%%%%%%%
%%%%%%%%%%%
\begin{equation}\label{eq:xsec}
\frac{d\sigma}{dT}(E_{\nu},T) = \frac{G_{F}^{2}}{8\pi} Q_{W}^2 \times M\left(2-\frac{T M}{E_{\nu}^{2}}\right)|f(q)|^2
\end{equation}  
%%%%%%%%%%%%%
and $Q_{W} = Z(4\mathrm{sin}^2\theta_{W} - 1)+N$.
%%%%%%%%%%%%%%%%%%%%%%%%%%
In Eq.~\ref{eq:xsec}, $M, N,$ and $Z$ are the mass, number of neutrons, and number of 
protons in the nucleus, 
respectively. Further, $E_{\nu}$ is the incident neutrino energy, $T$ is nuclear 
recoil energy, ($T_{\rm  max}(E_{\nu})=2E_{\nu}^{2}/(M+2E_{\nu})$), $G_{F}$ is the 
Fermi coupling constant, $\theta{_W}$ is the weak mixing angle, and $f(q)$ is the 
nuclear form factor for a momentum transfer of $q$. It can be noted that Eq.~\ref{eq:xsec} is
applicable for all types of neutrinos and antineutrinos. For low energy neutrinos 
($E_{\nu}<$ 50 MeV), the momentum transfer is very small such that $q^2R^2<$1, where 
R is the radius of the nucleus, which leads $f(q) \sim 1$. At small momentum transfers, the 
scattering amplitude from individual nucleons is in phase and adds coherently, 
which leads to the increase of the cross-section. The weak mixing angle $\mathrm{sin}^{2}\theta_{W}$ has been measured to be 0.23867 $\pm$ 
0.00016 
$\sim$1/4~\cite{Erler:2004in}. Then, the contribution from the proton is suppressed, that leads the cross-section proportional 
to $N^2$. Therefore, it is possible to detect neutrinos via the CE$\nu$NS process with kg-size detectors due to high event rates.
Further, it can be observed that the recoil energy is inversely proportional to the mass number of the target.
At a  given $E_{\nu}$, although the cross section can be enhanced by choosing a heavier
nucleus, the measurable recoil energy is lowered at the same time. Therefore, it is required 
to consider the  target materials with intermediate atomic numbers that is a good compromise
 between the experimental observable that corresponds the detection rate and detection possibility.
  It has been previously noted that measurement of low nuclear recoil energy
 of few keV presents a significant challenge, necessitating the minimization
  of low energy backgrounds in relevant measurements and the enhancement of accuracy in the CE$\nu$NS
  experimental approach.
  It is observed from Eq.~\ref{eq:xsec} that the cross section depends on
 weak mixing angle which  provides a complimentary measurement at low momentum transfer
 by measuring the signal due to low energy of recoil nuclei. Further, the measurement of
neutrino magnetic moment and other physics aspects can be addressed 
using the CE$\nu$NS process, which is discussed in the following subsections. 
%%%%%%%%%%%%%%%%%%%%%% Fig.1 %%%%%%%%%%%%%%%%%%%%%%%%%%%%%%%%%
%%%%%%%%%%%%%%%%%%%%%%%%%%%
\begin{figure}[t]
\includegraphics[width=0.4\textwidth]{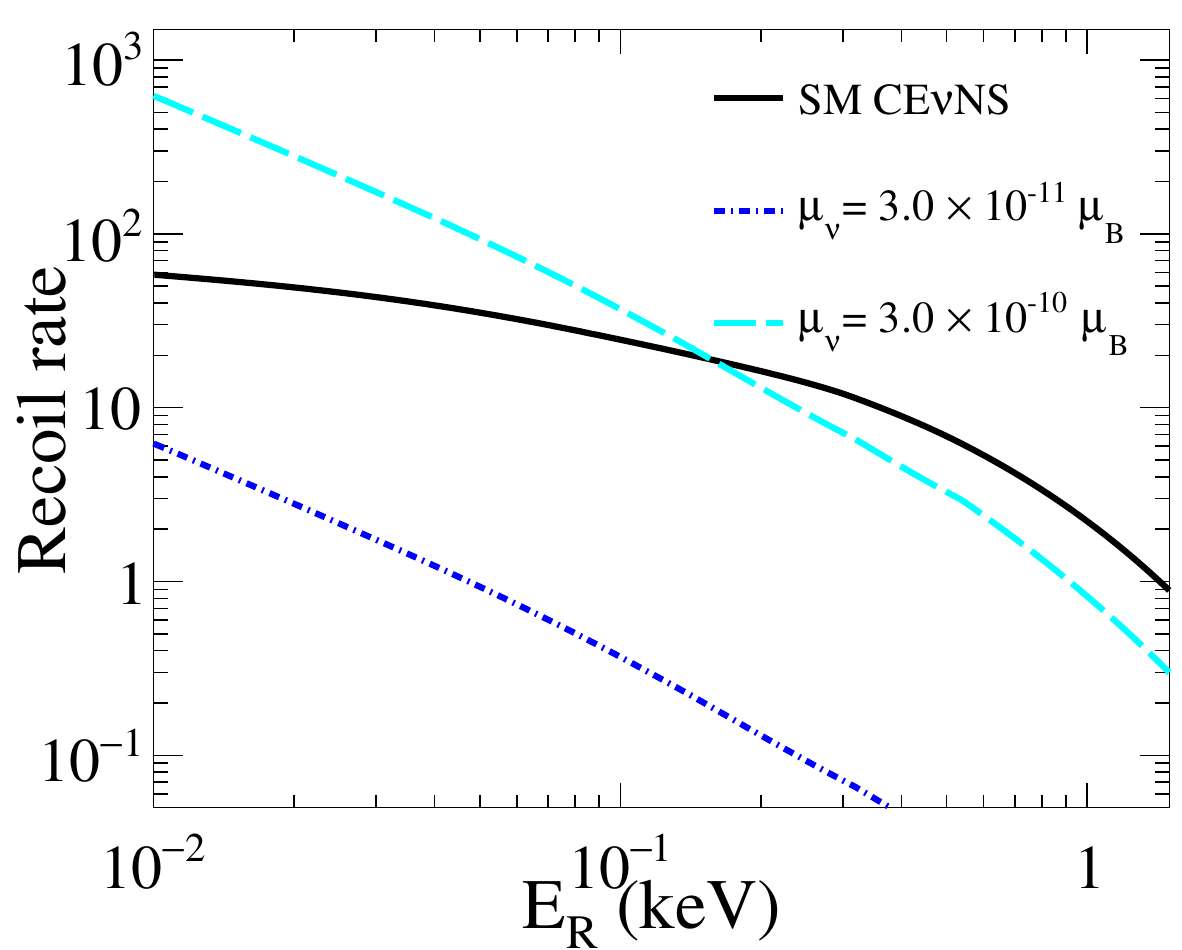}
\caption{ \label{fig:recoilMM} Recoil event rate per day as a function 
of nuclear energy at different values of neutrino magnetic moment for a sapphire 
detector of mass 10 kg is placed at a 4 m from the Apsara-U reactor core.} 
\end{figure}
%%%%%%%%%%%%%%%%%%%%%%
%%%%%%%%%%%%%%%%%%%%%%%%%%%
\begin{figure*}[t]
\centering
\includegraphics[width=0.4\textwidth]{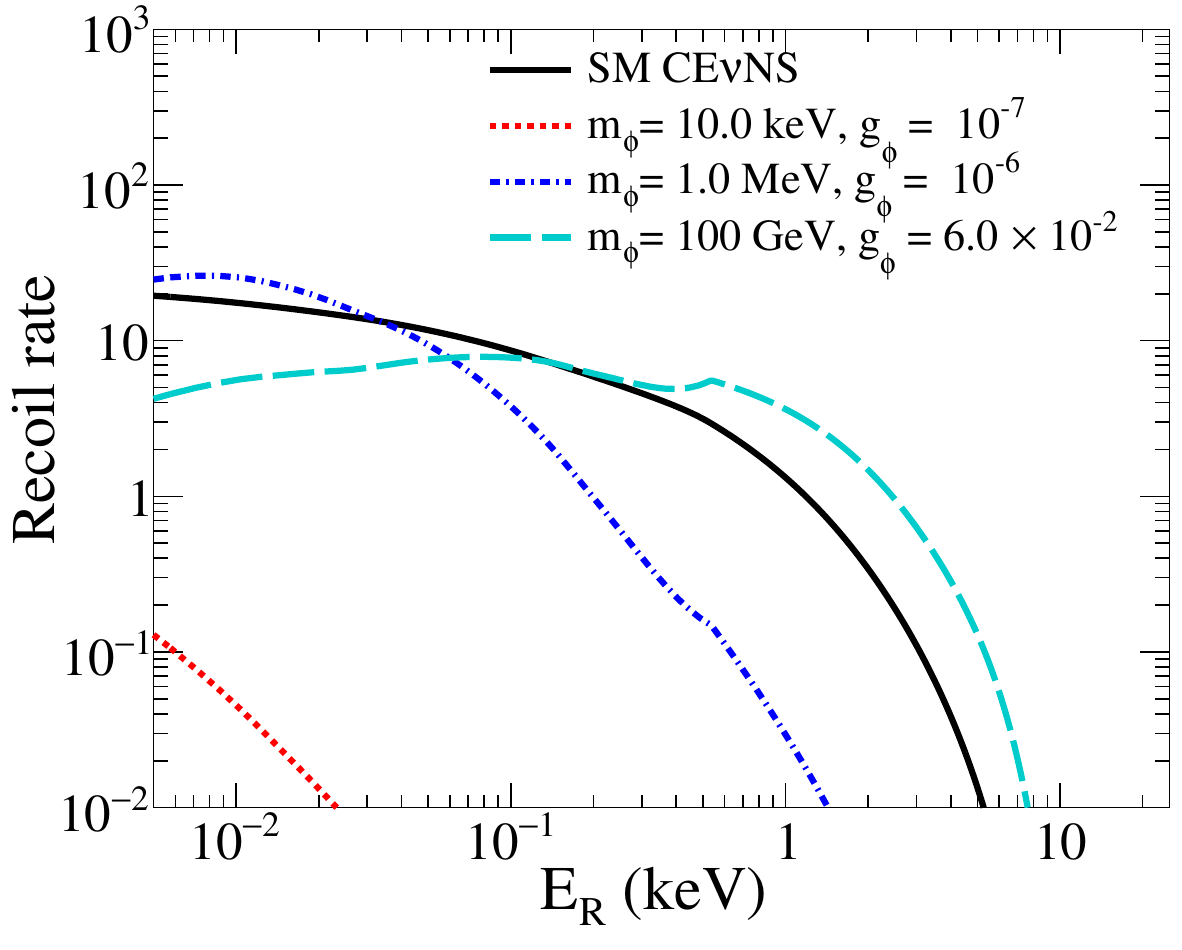}
\includegraphics[width=0.4\textwidth]{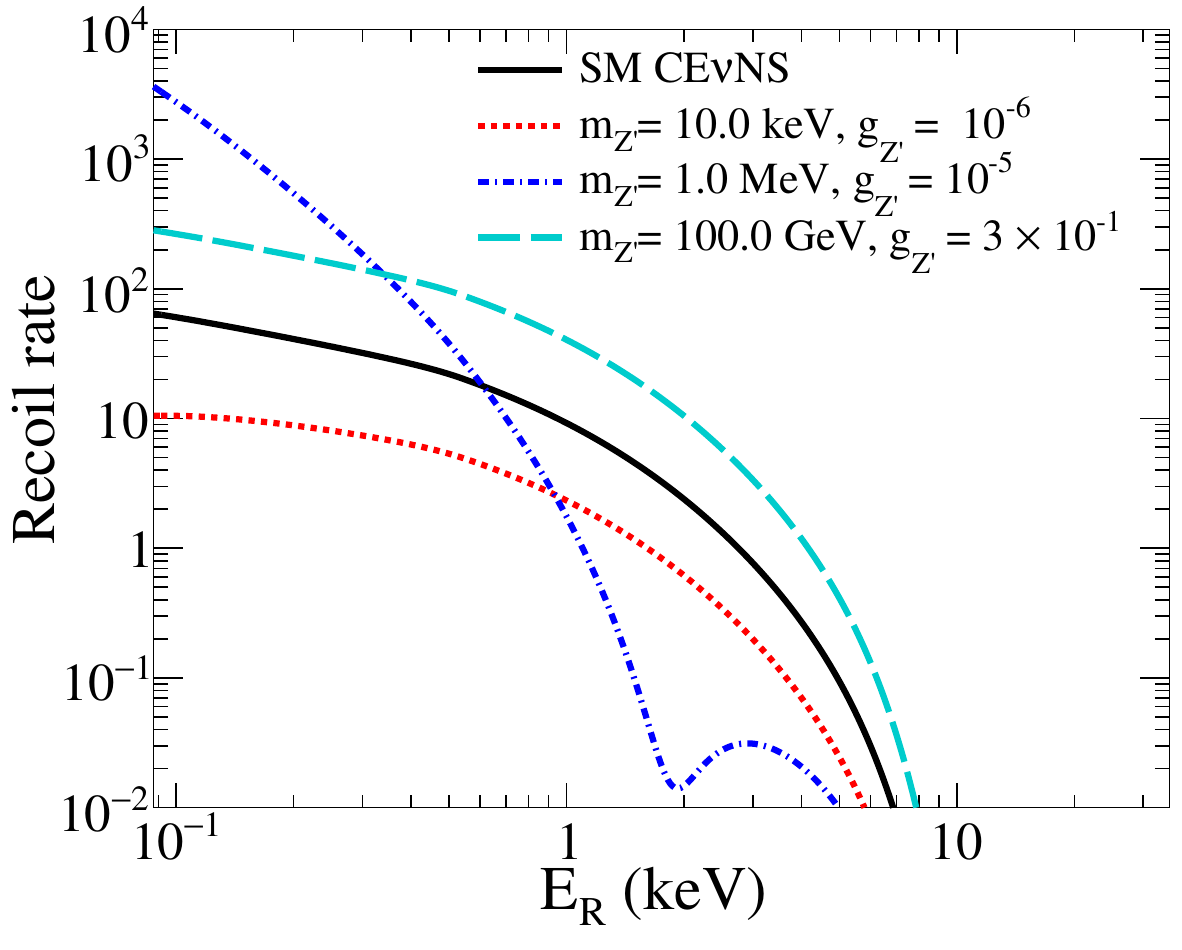}
\caption{ \label{fig:RRScalar} Recoil event rate as a function 
of nuclear energy due to exchange of scalar(left panel) and vector(right panel) mediators of different masses.
The sapphire  detector of mass 10 kg is placed at a 4 m from the Apsara-U reactor core.} 
\end{figure*}
%%%%%%%%%%%%%%%%%%%%%%
%%%%%%%%%%%%%%%%%%%%%%%%%%%%%%%%%%%%%%%%%%%%%%%%%%
%%%%%%%%%%%%%%%%%%%%%%%
\subsection{The CE$\nu$NS process due to the magnetic moment of neutrino}
%%%%%%%%%%%%%%%%%%%%%%%
%%%%%%%%%%%%%%%%%%%%%%%%%%%%%%%%%%%%%%%%%%%%%%%%%% 
In minimal extensions of the SM with three right-handed neutrinos, the
magnetic moment $\mu_{\nu}$ is predicted to be non-zero with 
values ranging from less than $10^{-14}$ to $10^{-19} \mu_{B}$, where $\mu_{B}$ represents the
Bohr magneton. The neutrino magnetic moment is a result of loop-level radiative correction~\cite{Lee}. 
In contrast, theories extending beyond  the minimal extended SM,
the neutrino magnetic moment could be on the order of $10^{-(10-12)}  \mu_{B}$  for
Majorana neutrinos, while for Dirac neutrinos, it  can
not exceed $10^{-14} \mu_{B}$. Hence, the observation of
neutrino magnetic moment can provide insight into the  nature of neutrinos.
The Super-Kamiokande Collaboration has put an upper limit of 3.6 $\times$ 10$^{-10}$ $\mu_{B}$ 
at a 90$\%$ confidence level (C.L.) using  solar neutrino spectra above 5-MeV.
It has been further improved to a limit of  1.1 $\times$10$^{-10}$ $\mu_{B}$ (90$\%$
C.L.) considering  additional information from other solar neutrino
and KamLAND experiments~\cite{Super-Kamiokande:2004wqk}.
The BOREXINO group has placed a constraint on  the magnetic moment of neutrinos using solar neutrinos,
 establishing an upper limit of the effective nuclear magnetic moment as 
$\mu_{\nu} < 2.8 \times 10^{-11}$ at a 90$\%$ C.L.~\cite{Borexino:2017fbd}. 
The best magnetic moment limit from the reactor antineutrinos based GEMMA
 experiment is 2.9 $\times$10$^{-11}$ $\mu_{B}$ (90$\%$
C.L.)~\cite{Beda:2013mta}.
Various groups 
have also placed constraints on the neutrino magnetic moment using neutrinos 
from different sources~\cite{TEXONO:2006xds,LSND:2001akn,Allen}.

The neutrino-nucleus scattering cross section modifies in the presence of 
a neutrino magnetic moment. In the simplified model framework, the scattering cross section is given by
%%%%%%%%%%%%%%%%%
%%%%%%%%%%%
\begin{equation}\label{eq:magxsec}
%\begin{split}
\frac{d\sigma^{mag}}{dT}(E_{\nu},T)  = \frac{\pi\alpha^{2}\mu_{\nu}^{2}Z^{2}}{m_{e}^2}
\left[\frac{1}{T}-\frac{1}{E_{\nu}}+\frac{T}{4E_{\nu}^2}\right] f(q)^2
%\end{split}
\end{equation}  
%%%%%%%%%%%%%
In the above equation, $\alpha$ is the fine structure constant, $\mu_{\nu}$ is the magnetic
moment of the neutrino, $m_e$ is the mass of the electron.
The recoil rate due to neutrino-nucleon magnetic scattering depends upon $1/T$ whereas 
its strength is controlled by the size of the effective neutrino magnetic moment
in units of Bohr magnetons $\mu_B$. Experimentally, a signature of a nonzero neutrino magnetic 
moment can be observed via distortion of the recoil spectrum of coherently scattered nuclei. 
The recoil event rate as a function of nuclear recoil energy is shown in Fig.~\ref{fig:recoilMM}
at various values of neutrino magnetic moment for 10 kg sapphire detector placed at a 4 m 
distance from the Apsara-U reactor core for an exposure of 1 year. It can be observed that at lower recoil energy
the recoil rate due to the SM CE$\nu$NS process is much flatter than the recoil rate from the electromagnetic interaction of neutrinos
due to their magnetic moment. The later rate increases  with increasing the magnetic moment.
Further, it is observed that at  the lower recoil energy, the more considerable increase in the
neutrino magnetic moment effect with respect to the SM CE$\nu$NS process.
So the detector with a lower threshold can differentiate the neutrino magnetic moment effect 
against the standard CE$\nu$NS process.
%%%%%%%%%%%%%%%%%%%%%%%%%%%%%%%%%%%%%%%%%%%%%%%%%%%%%
%%%%%%%%%%%%%%%%%%%%%%%%%%%%%%%%%%
\begin{table*}[t]
 \begin{center}
\caption{\label{tab:countRate}{Expected events rate in a sapphire detector for various reactor 
core to  the detector distance and thermal power. }}
\begin{tabular}{ ccc}
\hline
    Reactors name & Distance from core (m) & Counts/day/kg \\
\hline
   Apsara-U & 4.0 & 0.42 \\
    Dhruva & 10.0 & 3.25  \\
 PFBR & 25.0 & 5.26  \\ 
 VVER & 30.0 &  10.85  \\
   \hline
\end{tabular}
\end{center}
\end{table*}
%%%%%%%%%%%%%%%%%%%%%%%%%%%%%%%%%%%%%%%%%%%%%%%%%%%%%
%%%%%%%%%%%%%%%%%%%%%%%%%%%%%%%%%%%%
%%%%%%%%%%%%%%%%%%%%%%%%%%%%%
\subsection{The CE$\nu$NS process due to the exchange of  massive mediators}
%%%%%%%%%%%%%%%%%%%%%%%%%%%%%%%%%%%%%%
%%%%%%%%%%%%%%%%%%%%%
A new scalar particle $\phi$ can participate in the CE$\nu$NS process, which mediates an interaction
 between neutrinos and quarks. This results the modification of nuclear recoil energy spectrum, 
both for a light scalar as well as a heavy one. The interaction Lagrangian is given by
\begin{align}
\label{eq:L_scalar}
\mathcal{L}_{\phi} =  \phi \left[ g_\nu \overline{\nu_{R}}\nu_{L} + g_\nu^* \overline{\nu_{L}}\nu_{R}  + g_\ell \overline{\ell} \ell + g_u \overline{u} u + g_d \overline{d} d\right]\,
\end{align}
where $\nu_{L,R}$, are left- and right-handed neutrinos, $\ell = e,\,\mu~$, and $\tau$ are 
charged leptons, and $u$, $d$ are up- and down-type quarks.
It is noted here that the exchange of this new scalar
 mediator does not interfere with the SM $Z$-exchange. Then the modified SM CE$\nu$NS cross section
 due to $\phi$ exchange is expressed as~\cite{Cerdeno:2016sfi, Farzan:2018gtr}
%%%%%%%%%%%%%%%%%%%%%%%%%%%%%%%%%%%%%%%%%%%%%%%%%
%%%%%%%%%%%%%%%%%%%%%
\begin{equation}\label{eq:SMscalar}
\frac{\mathrm{d}\sigma_{\phi}}{\mathrm{d}T} = \frac{G_F^2}{4\pi} Q_{\phi}^2 \left(\frac{2M T}{E_{\nu}^{2}}\right) m_{N} F^{2}(q)\,.
\end{equation}
%%%%%%%%%%%%%%%%%%%%%%%%%%
In Eq.~\ref{eq:SMscalar} presented above, the scalar mediator's mass is denoted as $m_\phi$, the 
nuclear charge due to $\phi$ exchange is represented by $Q_\phi$, and the momentum transfer 
is calculated as $q = \sqrt{2 M T}$. The determination of the nucleus's `scalar charge' 
is approximated based on the quark couplings $g_q$ and the quarks contributions to the nucleon.
With the consideration for universal couplings to all quarks, an
 approximate formula is mentioned in Ref.~\cite{Farzan:2018gtr} that is expressed as
%%%%%%%%%%%%%%%%%%%
\begin{align}
Q_{\phi} = \frac{(15.1 \,Z + 14\, N)g_\phi^{2}}{\sqrt{2}G_{F}(2MT+m_{\phi}^2)} ,
\end{align}
%%%%%%%%%%%%%%%%%%%%%%%%%%%%%%%%%%
%%%%%%%%%%%%%%%%%%%%%% Fig. %%%%%%%%%%%%%%%%%%%%%%%%%%%%%%%%%
where $g_\phi^{2} = g_{\nu}g_{q}$, with $g_{\nu}$ is the neutrino coupling and $g_{q}$ is the common coupling to quarks.
  Left panel of Fig.~\ref{fig:RRScalar} illustrates the impact on the nuclear recoil rate caused by a  scalar mediator,
 featuring various masses and couplings. Solid line shows the SM  CE$\nu$NS rate  and the contribution
  due to  the exchange of scalars with different masses is shown in dashed(dashed-dotted) lines.
It has been observed that for light mediators with masses less than or on the order of MeV, 
the interaction exhibits an effectively long-range behavior. This results in a recoil spectrum 
that rapidly decreases with momentum transfer and follows a relationship of $q^{-4}$ ($\sim T^{-2}$).
 On the other hand, for heavier mediators, the spectrum becomes peaked, with the cross section 
 scaling linearly with the recoil energy ($T$) at low energies. However, at high energies, the 
 coherence is lost, leading to a cutoff in the spectrum. In a study by Farzan et al.~\cite{Farzan:2018gtr}, 
 it is mentioned that if the mass of the new scalar is similar to the energy of the neutrinos, it should 
 also be feasible to measure the mass of the mediator ($m_\phi$).
  Similarly, the SM CE$\nu$NS cross section can be modified due to the presence of a new massive vector
 mediator that couples to the SM neutrinos and quarks. In the present study we have considered  a vector 
 neutral current neutrino non-standard interaction with quarks that exhibits a coherent nuclear 
 enhancement due to contributions from $Z$ and $Z^{'}$ exchange, leading to its typical dominance.  
 Further, it is assumed that the exchange of this new vector mediator does not interfere
with the SM Z-exchange. 
 Due to the exchange of this vector mediator, $Q_{W}$ mentioned in Eq.~\ref{eq:xsec} is modified, which is given by~\cite{Billard:2018jnl}
 %%%%%%%%%%%
 \begin{equation}
  Q_{SM+Z^{'}} =  Q_{W} - \frac{\sqrt(2)}{G_{F}} \frac{Q_{Z^{'}}}{q^{2}+m^{2}_{Z^{'}}} 
 \end{equation} 
 %%%%%%%%%%%%%
 where $Q_{Z'} = (2Z + N )g_{u} g{_\nu} + (2N + Z)g_{d} g_{\nu}$ which leads to a change in 
 differential cross section. Right panel of Fig.~\ref{fig:RRScalar} shows the effect on the
  nuclear recoil rate caused by a  vector mediator  of different masses and couplings.
   It is found that recoil rate increases at low energy due 
 to the exchange of vector mediators with an increase in mass.
 %%%%%%%%%%%%%%%%%%%%%%%%%%%%%%%
\begin{table*}[t]
 \begin{center}
\caption{\label{tab:sysUnce}{List of systematic uncertainties  
considered in the analysis. }}
\begin{tabular}{ cc}
\hline
    Uncertainties  & Contribution($\%$)\\
\hline
    Total
neutrino flux, number of target atoms, and detector efficiency & 5.0  \\
    Energy
calibration & 1.0  \\
Nonlinear energy response & 1.0   \\ 
\hline
\end{tabular}
\end{center}
\end{table*}
%%%%%%%%%%%%%%%%%%%%%%%%%%%%%%%%
 %%%%%%%%%%%%%%%%%%%%%%%%%%%%%%%%%
%%%%%%%%%%%%%%%%%%
\section{EVENT RATE IN A DETECTOR}
\label{sec:expect}
%%%%%%%%%%%%%%%%%%%%%%%%
A high CE$\nu$NS reaction cross section per unit detector mass is advantageous for 
detectors weighing in the kilogram range. The predicted events rate due to
 CE$\nu$NS can be calculated as
%%%%%%%%%%%%%%%%%%%%%
\begin{equation}\label{evt}
\begin{split}
 N^{\rm SM}_{\rm events} &=\epsilon t\lambda_{0} \frac{M_{\text{detector}}}
 {A}\int_{E_{\nu \rm min}}^{E_{\nu \rm max}}\lambda(E_{\nu})dE_{\nu}\\
&\int_{T_{\rm min}}^{T_{\rm max}(E_{\nu})}\left(\frac{d\sigma}
 {dT}\right) dT ,
\end{split}
\end{equation}
%%%%%%%%%%%%%%%%%
\noindent where $M_{\text{detector}}$ is the mass of the detector, $t$ is the 
time duration of data taking, $\lambda_{0}$ is the total neutrino flux, 
$\lambda(E_{\nu})$ is the neutrino spectrum and $\epsilon$ is the efficiency of the detector. 
In the present study, the Hubber-Muller 
model~\cite{Huber:2011wv,Mueller:2011nm}  parameterization has been considered for 
\antinue s produce from beta decay of fission fragments of  energy spectra above 2.0 MeV. 
The low energy part of the \antinue s spectra 
is considered from Ref.~\cite{Vogel:1989iv,TEXONO:2006xds}. 
The antineutrinos produced from the slow neutron capture by the $^{238}$U 
 have energy $<$2 MeV. We have considered the numerical data for this part of the 
 spectrum from Ref.~\cite{TEXONO:2006xds}.
The minimum recoil energy of the nuclei in a 
specific experimental setup is determined by the detector threshold. We have chosen to analyze neutrinos 
with a maximum energy of approximately 10.0 MeV 
since there are 
fewer neutrinos above this energy. 
The number of events expected in the detector due to various reactor power as well as reactor
core to the detector distance is listed in the Table.~\ref{tab:countRate}. Event rate has been
estimated assuming a detection energy threshold of 100 eV, 80$\%$ of 
the detection efficiency independent of nuclear recoil, 
90$\%$ fiducial volume of the detector, 70$\%$ 
reactor duty cycle and, for an exposure of  1 day. 
Studies have been conducted using a target mass of 10 kg to extract the detector sensitivity for 
various parameters related to the CE$\nu$NS process.It can be noted here that the remaining 
studies are conducted for an exposure period of one year.
%%%%%%%%%%%%%%%%%%%%%%%%%%%
\begin{figure*}[t]
\centering
\includegraphics[trim=4.2 0 0 0, clip,width=0.34\textwidth]{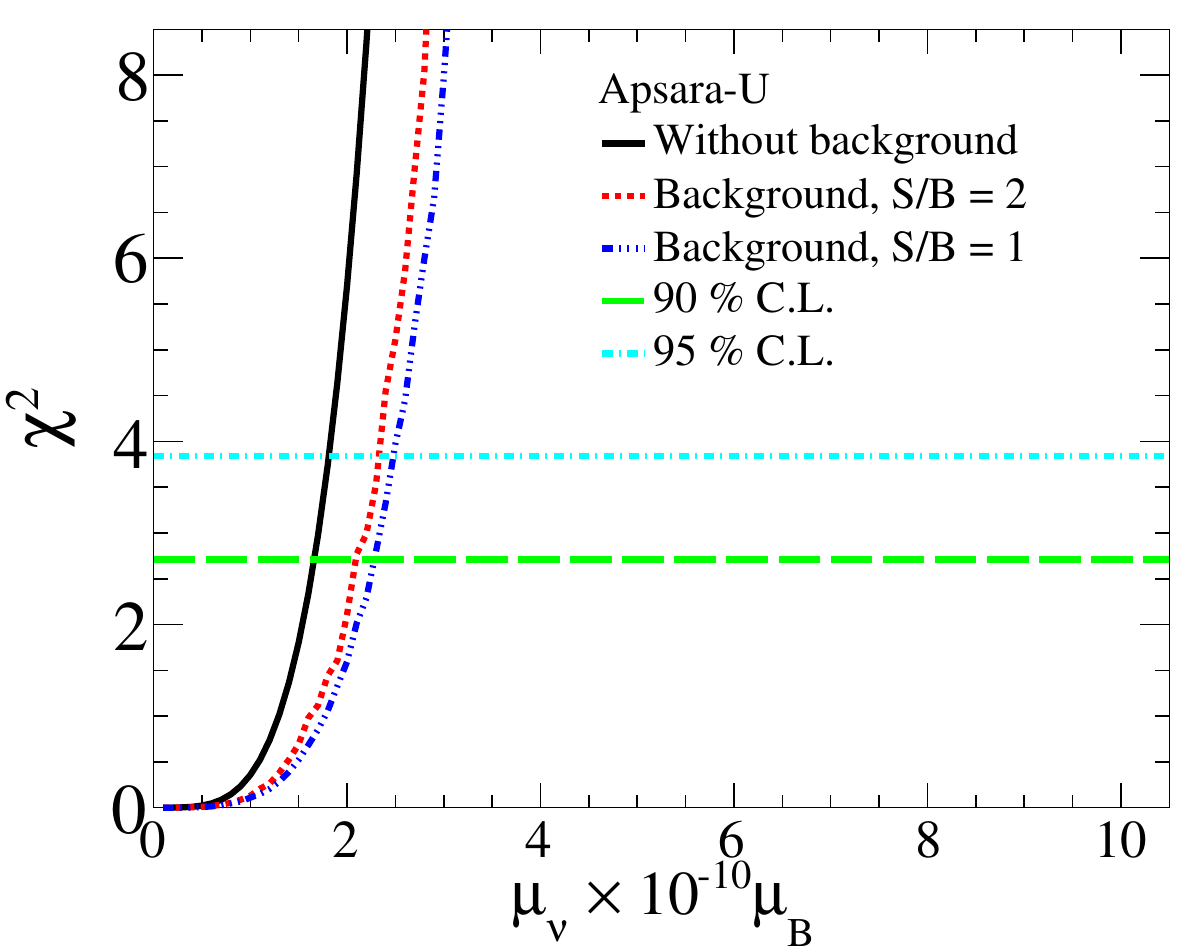}
\includegraphics[trim=4.2 0 0 0, clip,width=0.34\textwidth]{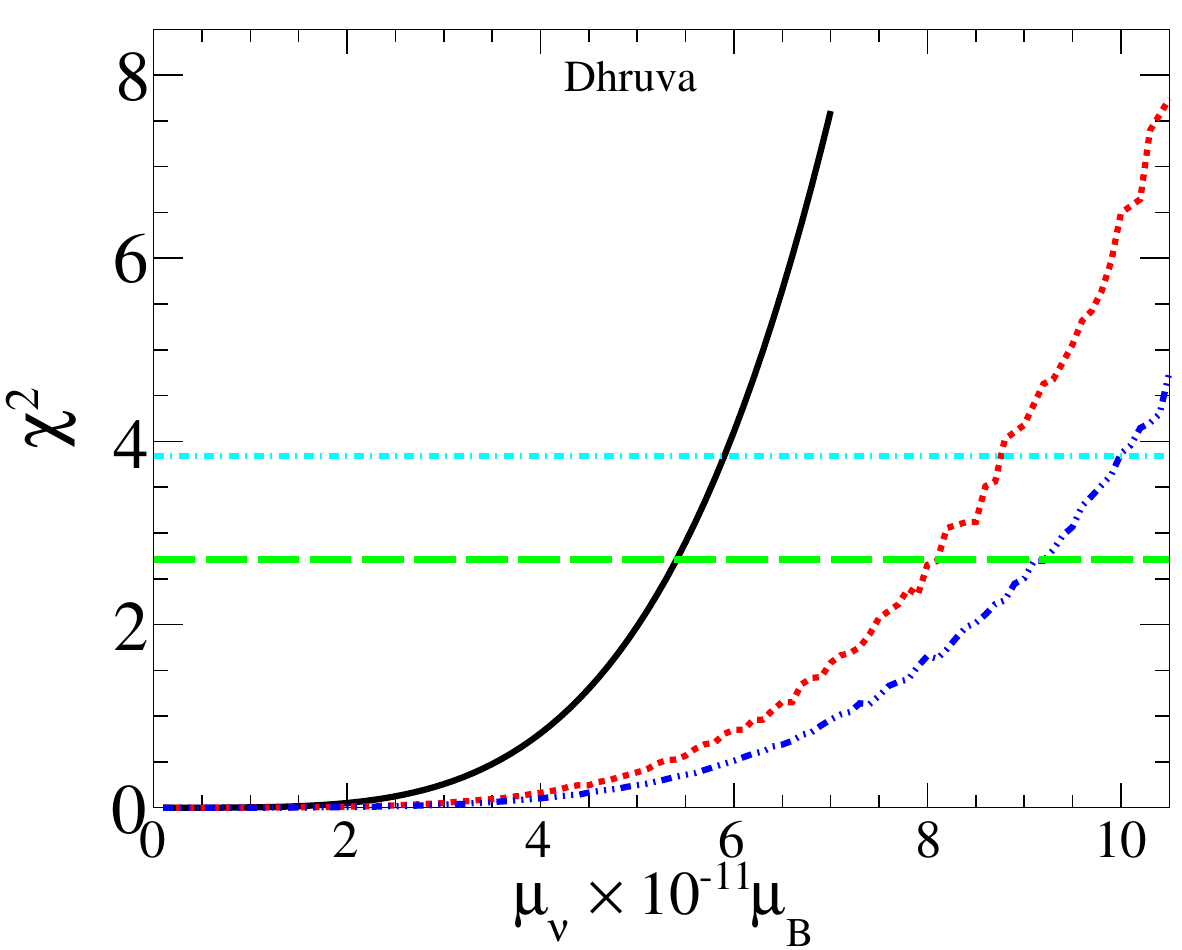}
\includegraphics[trim=4.2 0 0 0, clip,width=0.34\textwidth]{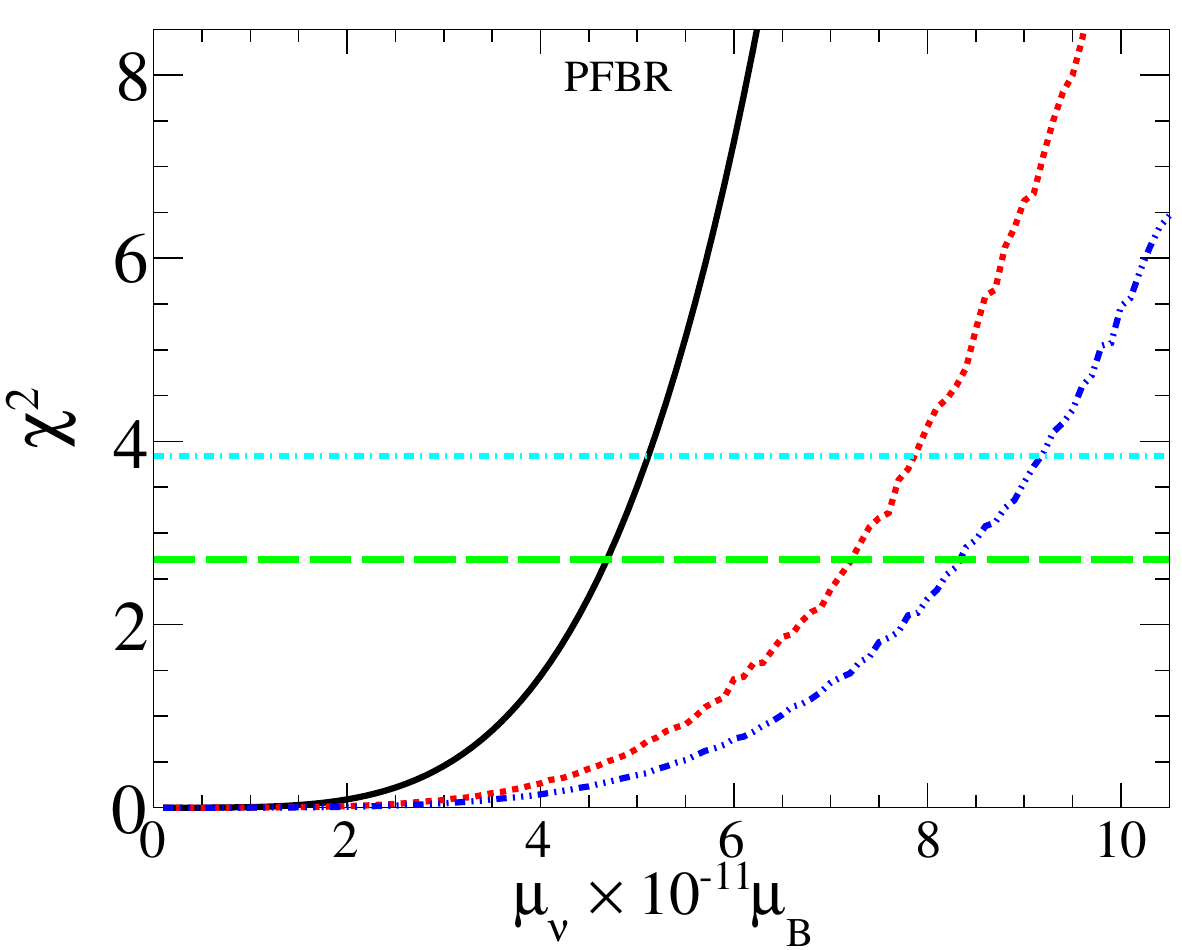}
\includegraphics[trim=4.2 0 0 0, clip,width=0.34\textwidth]{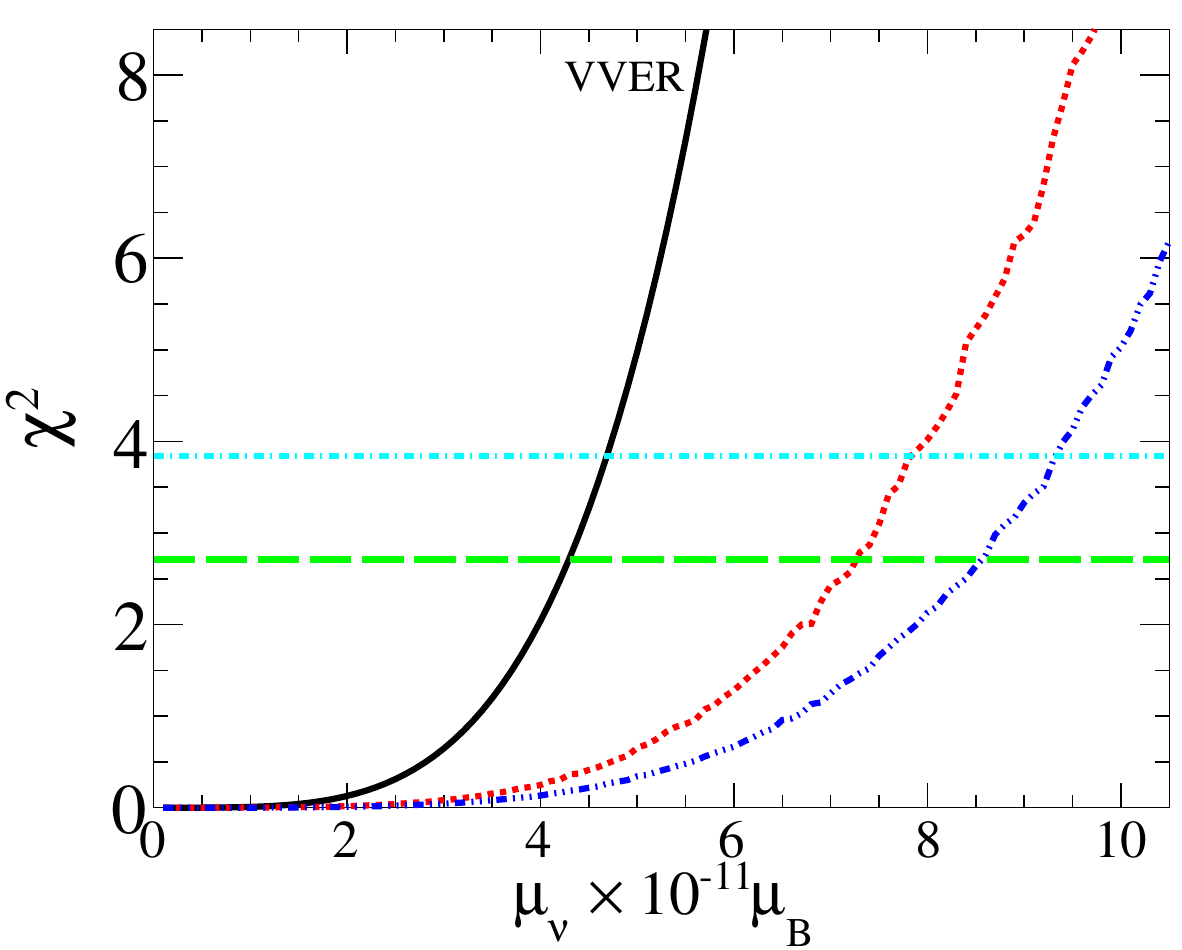}
\caption{ \label{fig:MagMom10kg} Comparison of the ICNSE detector sensitivity to the neutrino magnetic
 moment considering with and without background.} 
\end{figure*}
%%%%%%%%%%%%%%%%%%
%%%%%%%%%%%%%%%%%%%%%%%%%%%%%%%%%%%%%%%%%%%%%%%%%%%%
%%%%%%%%%%%%%%%%%%%%%%%%%%%%%%%%%%%%%%%%%%%%
\section{Simulation Method and Extraction of the Detector Sensitivity }
\label{sec:simul}              %
%%%%%%%%%%%%%%%%%%%%%%%%%%%%%
The present study has been carried out to find the potential of the ICNSE
 detector on different physics parameters by using \antinue s produced from various 
types of reactor facilities that is mentioned in Sec.~\ref{sec:reactor}. The number 
of neutrinos produced from the reactor depends  both on the thermal power as well as 
on fuel compositions. Electron antineutrinos flux produced from 
the reactor has different energy dependent for various isotopes. The parametrization for $\antinue$ flux assumed in 
the present analysis is mentioned in Ref.~\cite{Behera:2023llq}.
The number of events expected in the detector has been evaluated by knowing the energy dependent  
flux, cross-section,  the detector efficiency, fiducial volume of the detector and the reactor duty cycle . 
 Both the production point of 
neutrinos inside the reactor core and the interaction point in the detector are 
generated using a Monte-Carlo method.
The sensitivity is evaluated by assuming that a given
experiment searching for CE$\nu$NS events will measure
exactly the SM expectation; thus any deviation is understood as a signature of new physics.
 For this purpose, a statistical analysis between the 
predicted and expected event distribution obtained from simulation is carried out in order 
to quantify the sensitivity of the detector for a given exposure. The sensitivity 
of the detector to various  parameters is extracted by 
estimating the $\chi^2$.
 %%%%%%%%%%%%%%%%%%%%%%%%%%%
\begin{figure*}[t]
\centering
\includegraphics[width=0.34\textwidth]{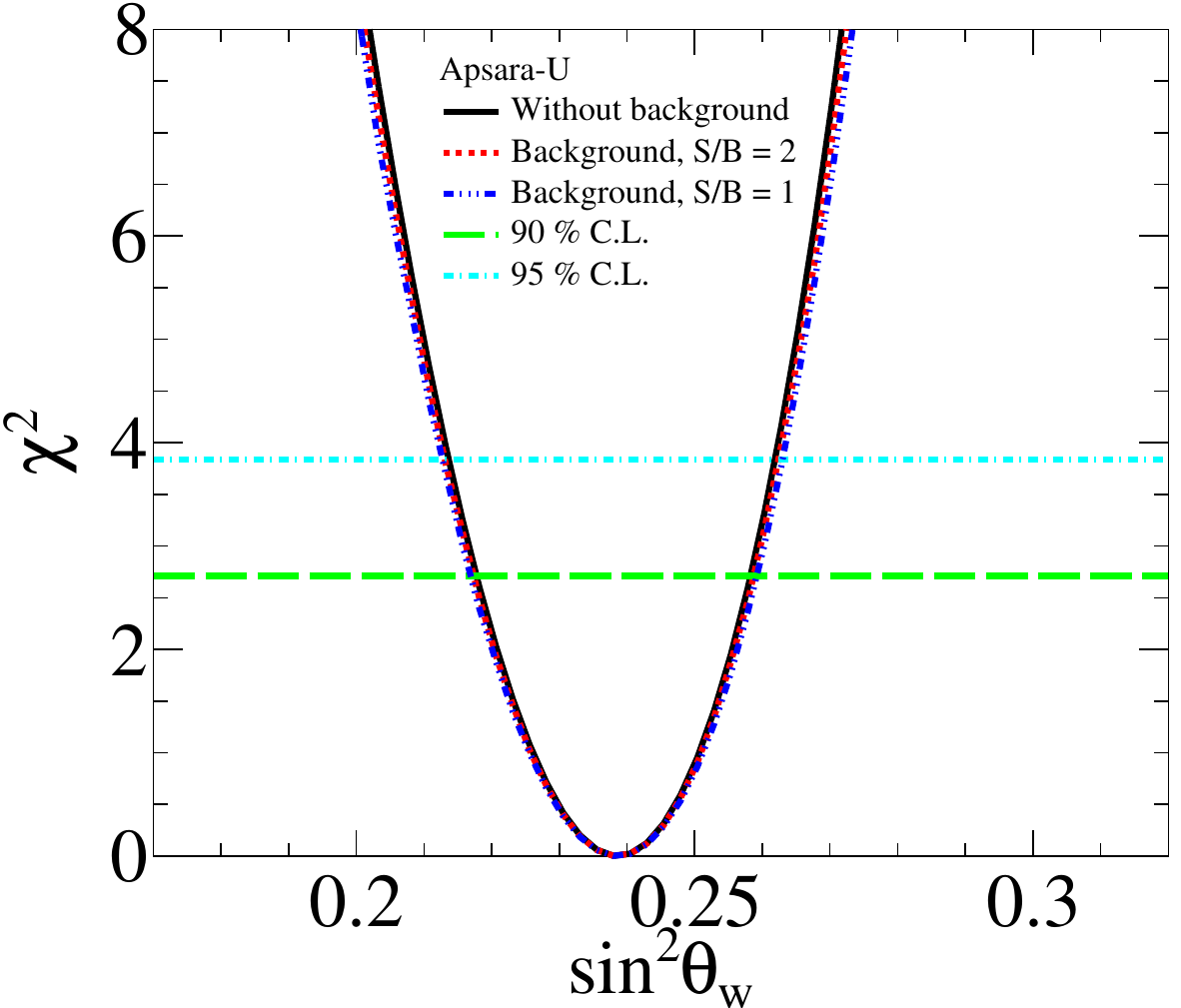}
\includegraphics[width=0.34\textwidth]{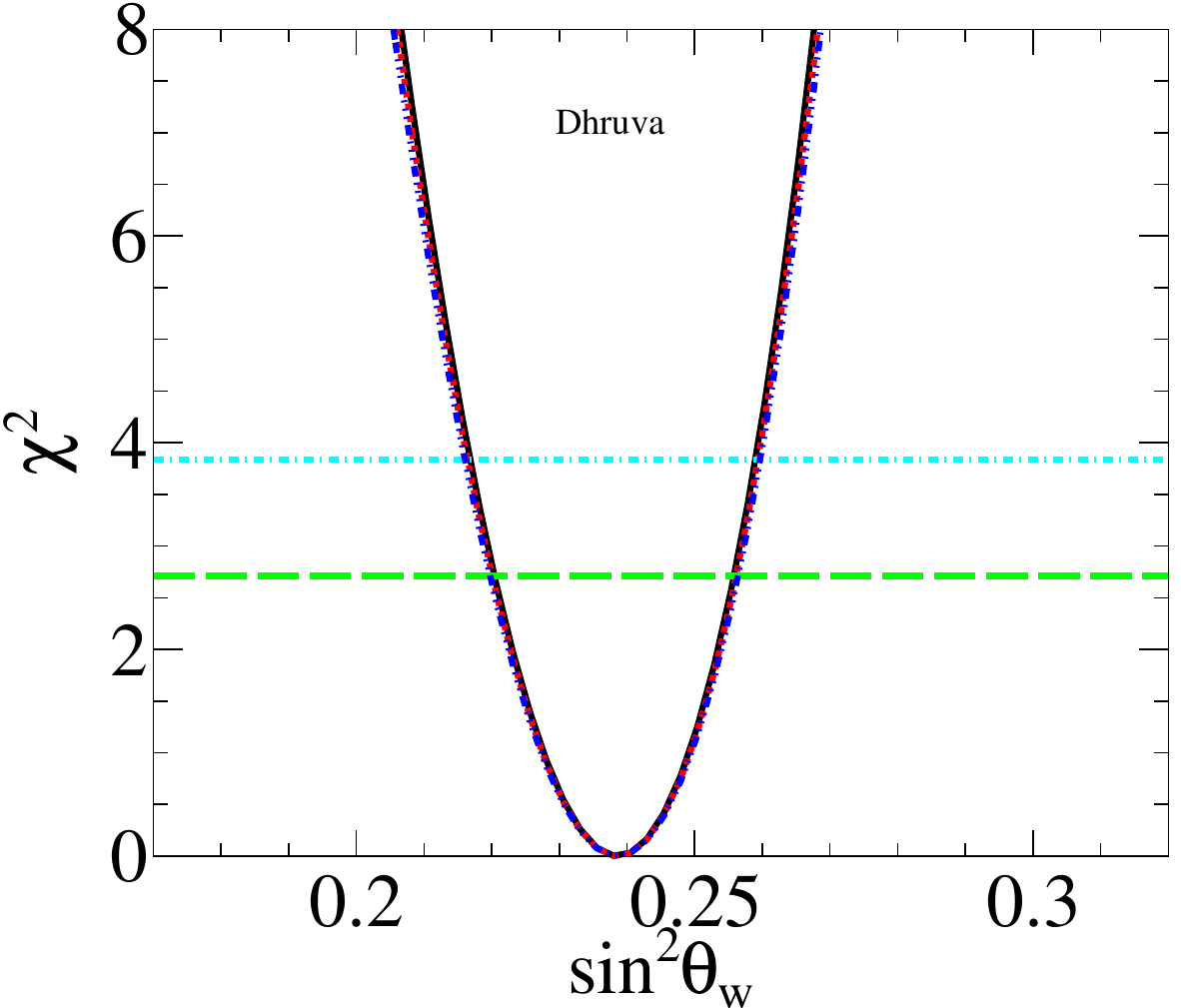}
\includegraphics[width=0.34\textwidth]{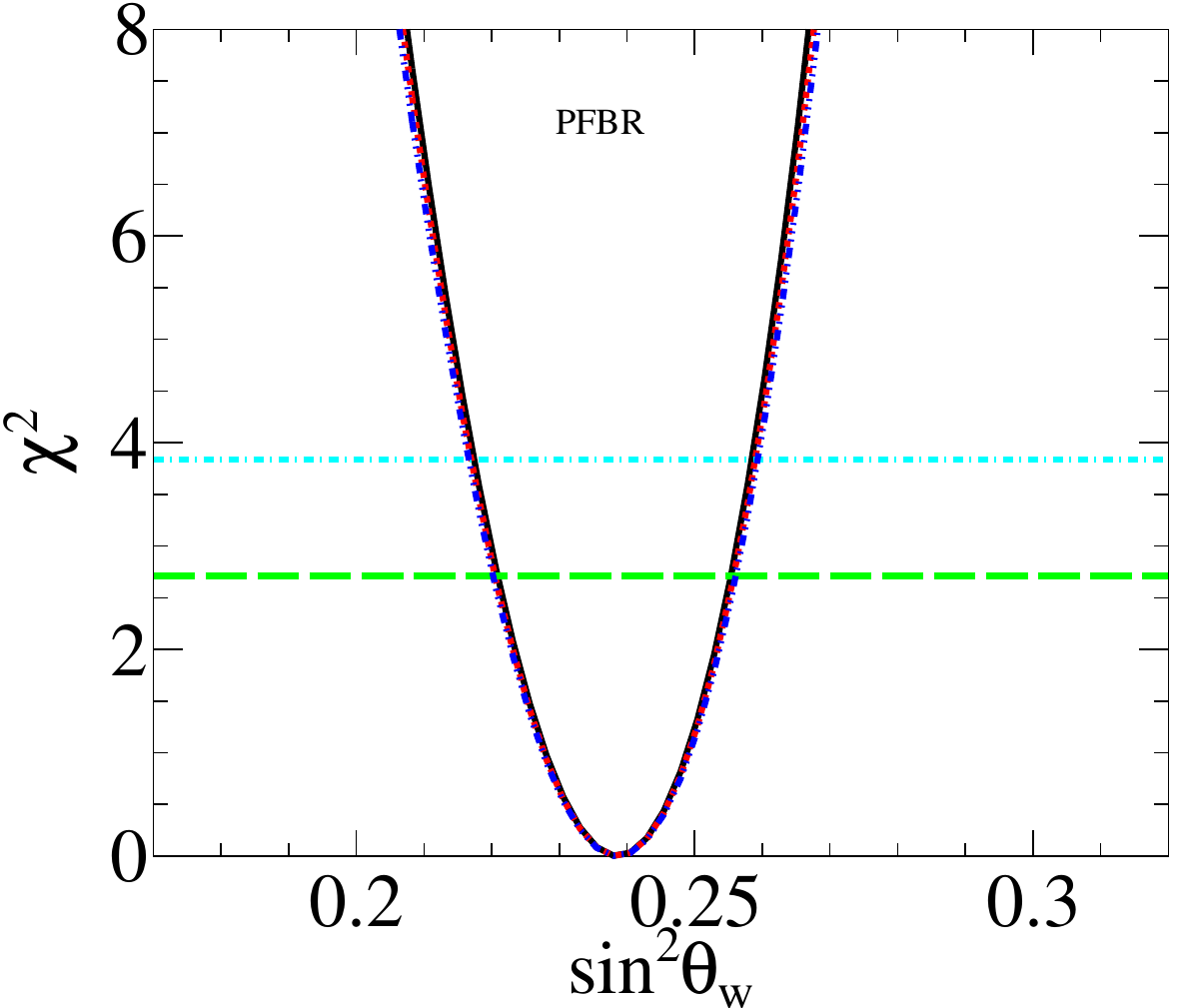}
\includegraphics[width=0.34\textwidth]{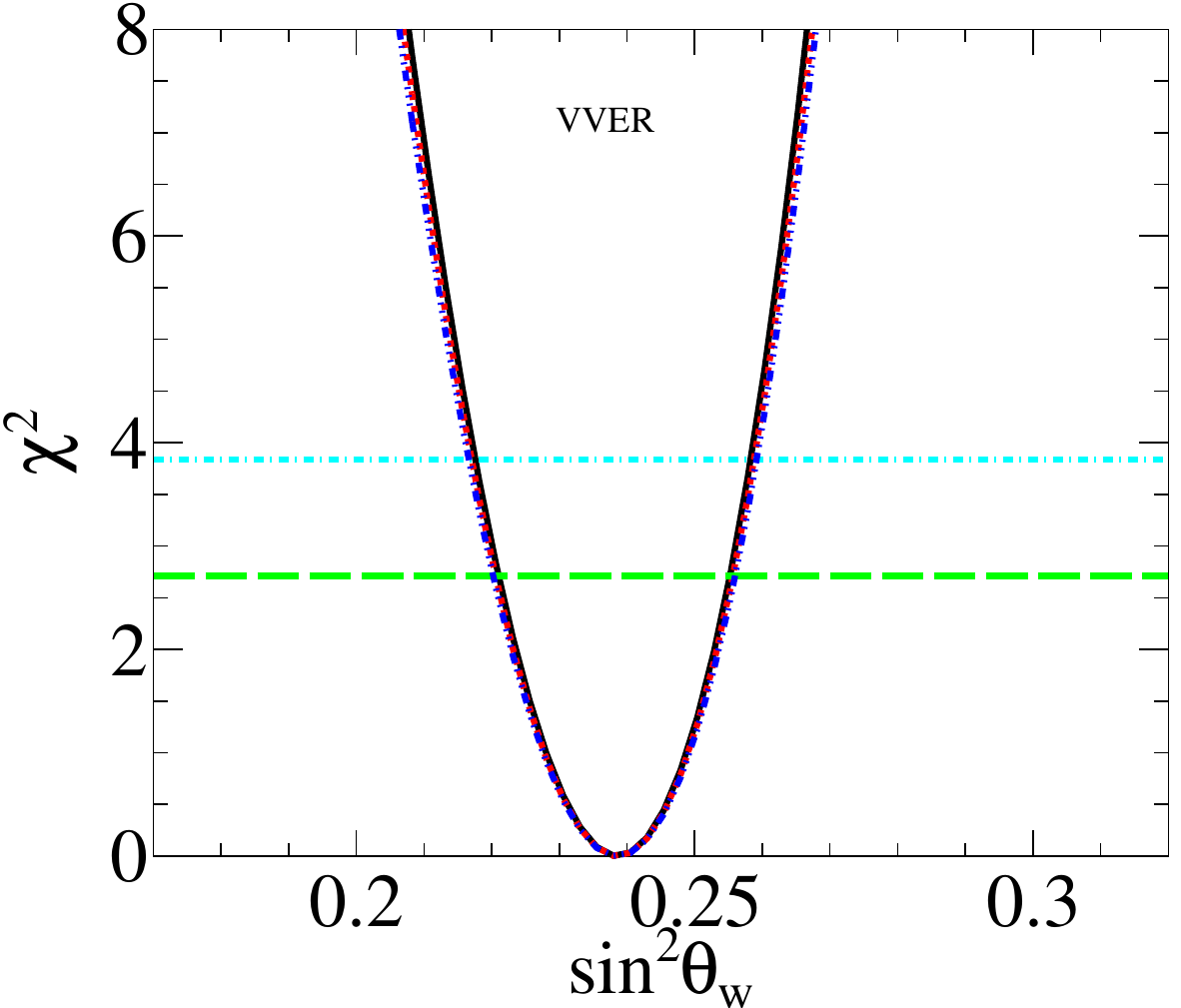}
\caption{ \label{fig:weakAng} Sensitivity of the detector to the weak mixing angle considering 
Apsara-U, Dhruva, PFBR, and VVER reactors as antineutrinos sources.} 
\end{figure*}
%%%%%%%%%%%%%%%%%%%%%%%%%%%%%%%%%
The definition of the $\chi^{2}$ is taken from Ref.~\cite{Lindner:2016wff} and 
given as
%%%%%%%%%%%%%%%%%%%%%%%%%%%%%
\begin{equation}
\chi^{2}=\frac{\xi^{2}}{\sigma_{\xi}^{2}}+\sum_{T\thinspace{\rm bins}}\frac{[(1+\xi)N_{n}^{th}(\xi)-N_{n}^{ex}]^{2}}{\sigma_{{\rm stat},n}^{2}+\sigma_{{\rm sys},n}^{2}},\label{eq:chi1}
\end{equation}
%%%%%%%%%%%%%%%%%%%%
where $\xi$ denotes the pull parameter with uncertainty $\sigma_{\xi}$.  
In Eq.~\ref{eq:chi1}, $N^{ex}$, $N^{th}$ are representing the number 
of events obtained from the simulations with the  deviation from the SM CE$\nu$NS cross section 
(considered as measured) and with consideration of the SM CE$\nu$NS cross section 
( considered as theoretically predicted) events, respectively.
The procedure for estimating theoretically 
predicted events $N_{n}^{th}$  with consideration of the reactor as well as the detector related parameters is
 mentioned in Ref.\cite{Behera:2023llq}. 
 In both types of simulated events, the detector
response such as resolution and efficiency are incorporated.
The procedure for the detector response incorporation is mentioned in Ref.~\cite{Behera:2023llq}.
 The statistical uncertainty $\sigma_{{\rm stat},n}$
and the systematic uncertainty $\sigma_{{\rm sys},n}$ of the event
number in the $n$-th recoil energy bin are given by
%%%%%%%%%%%%
\begin{equation}
\sigma_{{\rm stat},n}=\sqrt{N_{n}^{th}+N_{{\rm bkg},\thinspace n}}\,,\,\,\sigma_{{\rm sys},n}=\sigma_{f}(N_{n}^{th}+N_{{\rm bkg},\thinspace n})\
\end{equation}
%%%%%%%%%%%%%%
Here $N_{{\rm bkg},\thinspace n}$ is the number of background events.
We assume that $\sigma_{{\rm sys},n}$ is proportional
to the event number with a coefficient $\sigma_{f}$.
The $\chi^2$ is minimized with respect to pull 
variables $\xi$ and it is estimated by considering different sources of systematic 
uncertainties as mentioned in Table~\ref{tab:sysUnce}. It includes  normalization uncertainty which arises due to 
reactor total neutrino flux, the number of target atoms, and the detector efficiency, 
uncertainty due to the nonlinear energy response of the detector and, 
uncertainty in the energy calibration. So an overall $\sigma_{\xi}$ = 5$\%$ systematic uncertainty
has been considered.  

The measurement of  low energy of recoil nuclei from the CE$\nu$NS process faces a
significant challenge due to the presences of both natural and reactor backgrounds
such as gamma-rays and neutrons.   The sensitivity of the detector is influenced 
not only by the various types of backgrounds but also by the energy-dependent 
configuration of these backgrounds. At low recoil energy, two distinct background 
shapes are observed: the $1/T$ shape and flat-shaped backgrounds, as referenced
in Ref.~\cite{Bowen:2020unj}. The procedure for extracting the detector 
sensitivity due to the presence of different types of backgrounds
 is mentioned in~\cite{Behera:2023llq}. Similarly the systematic 
 uncertainty due to backgrounds is considered as $\sigma_{f}$ = 5$\%$.
 A signal-to-background ratio of 1.0 and 2.0 has been taken into 
 consideration to determine the sensitivity of the detector in the presence of background.
%%%%%%%%%%%%%%%%%%%%%%%%%%%%%%%%%%%%%%%%
\section{Results and Discussions}\label{sec:results}
%%%%%%%%%%%%%%%%%%%%%%%%%%%%%%%%%%%%%%%%%%%%%
The sensitivity of the detector to various physics parameters are computed considering
the procedure mentioned above.
Results based on this study are presented below.
%%%%%%%%%%%%%%%%%%%%
\subsection{Sensitive to the magnetic moment of neutrinos}
%%%%%%%%%%%%%%%%%%%%
%%%%%%%%%%%%%%%%%%%%%%%%%%%%%%%%%%%%%%%%%%%%%%%%%%%%%%%%%%%%%%
The detector's sensitivity to the neutrino magnetic moment has been determined by analyzing the simulated events.
The theoretically predicted events, which take into account the SM CE$\nu$NS cross section, are 
compared with the simulated measured events estimated using the cross section affected by neutrino magnetic moments.
It is found that the ICNSE detector has sensitivity to the interesting 
regimes of the neutrino magnetic moment. The detector sensitivity to the neutrino magnetic 
moment is shown in Fig.~\ref{fig:MagMom10kg} taking into account the different reactor
 power and the distance of the reactor core to the detector as previously 
 mentioned. In Fig.~\ref{fig:MagMom10kg}, solid and dotted(dashed-dotted) lines show
the detector sensitivity considering without and with background, respectively.
The results are compared with sensitivity at 90.0$\%$ and 95.0$\%$ C.L. 
It has been observed that the detector placed at the Apsara-U reactor can measure the neutrino magnetic moment
for $\mu_{B} \geq 1.66\times 10^{-10} \mu_{B}$ in the absence of background. 
The neutrino magnetic moment can be restricted even further 
 below the level of $\mu_{B} \sim 5.4\times 10^{-11}$,
 $\mu_{B} \sim 4.68\times 10^{-11}$, and $\mu_{B} \sim 4.29\times 10^{-11}$  
at 90 $\%$ C.L. using the ICNSE detector without having  background, placing at a reactor with high power such as 
Dhruva, PFBR, and VVER facilities, respectively.  The GEMMA experiment has put the strongest 
 constraint on the neutrino magnetic moment using reactor 
antineutrinos, which is $<$ 2.9 $\times 10^{-11} \mu_{B}$ (90$\%$ C.L.).  
 Furthermore, the detector's sensitivity is extracted based on the backgrounds present at the experimental site.
 In the present study, a signal-to-background ratio(S/B) of 2.0 and 1.0 are considered. 
 The detector sensitivity reduces with the inclusion of background,
  as demonstrated in Fig.~\ref{fig:MagMom10kg}. In the case of the detector
 placed at the Apsara-U reactor, it has been observed that the sensitivity 
 reduced by about 25.9 $\%$ and 37.3$\%$ considering a
  signal-to-background ratio of 2.0 and 1.0, respectively, at 90.0 $\%$ C.L.
 The sensitivity of the detector reduces by about 50.0 $\%$ and 53.2$\%$ for S/B = 2.0 by
  placing it at the Dhruva and PFBR reactor facilities, respectively. Additionaly,
  it has been bound that with increasing the 
systematic effect($\sigma_{f}$) 10 $\%$ from 5$\%$, the sensitivity of the 
detector has been reduced further by about 15.2$\%$
for S/B= 1.0 , considering the detector placed at 10 m from the Dhruva reactor core.
%%%%%%%%%%%%%%%%%%
%%%%%%%%%%%%%%%%%%%%%%%%%%%%%%%%%%%%%%%%%%%
\subsection{Sensitive to the weak mixing angle}
%%%%%%%%%%%%%%%%%%%%%%%%%%%%%%%%%%%%%%%%%%%
The weak mixing angle is one of the fundamental parameters of particle physics, which 
couples the electromagnetic and weak interactions. In the CE$\nu$NS process,  the cross-section is
proportional to the weak charge $Q_{W}^{2}$ as mentioned in Eq.~\ref{eq:xsec}. Hence, the weak 
mixing angle can be extracted from
the measured absolute cross section. In an experiment, it can be measured by observing a deviation 
from the SM prediction of the weak mixing angle. It has been measured precisely by the Z-pole 
experiment at high energy. However, at low energy measurements are carried out with less precision. 
There are several measurements carried out at low energy, such as Qweak~\cite{Qweak:2018tjf},  
and from the atomic parity violation experiments~\cite{ParticleDataGroup:2016lqr}. Further measurements by
 various group may improve the precision as mentioned in Refs.~\cite{Becker:2018ggl,Souder:2016xcn,MOLLER:2014iki}.
At low energy, the precision of weak mixing angle measurement may be improved by the CE$\nu$NS process. 
The weak mixing angle sensitivity of the ICNSE detector has been extracted considering 
reactors of different thermal powers and at different reactor cores to the detector
 distance. The sensitivity 
has been extracted by estimating the chi square between the simulated events with consideration of 
the SM weak mixing angle $sin^2\theta_W$ = 0.2386  and events generated by varying the weak mixing angle.
Figure~\ref{fig:weakAng} shows the possible sensitivity of the detector to weak mixing angle
 at 90$\%$ C.L. considering different reactors as neutrino sources.
The uncertainty expected due to measurement has also been extracted.
At 90$\%$ C.L., the width of the 
weak mixing angle $\delta sin^2\theta_W$ is estimated as $( S_{W^{max}}^{2} - S_{W^{min}}^{2})/2$ 
and the corresponding uncertainty is $\delta sin^2\theta_W/sin^2\theta_W$, where $S_{W^{max}}$ 
and $S_{W^{min}}$ are the upper and lower limits values at 90$\%$ C.L., respectively. 
It has been observed that expected uncertainties on
weak mixing angle measurements are about 8.59 $\%$,  7.33 $\%$, 7.33$\%$ and, 7.12$\%$ by placing the detector  at 
Apsara-U, Dhruva, PFBR, and VVER reactor facilities, respectively.
A larger uncertainty arises due to the fact that the detector has a low Z
 target material.  In order to improve the measurement
 sensitivity, it is necessary to have a highly intense neutrino source or a detector of larger mass,
which reduce the statistical error. 
 Further, it has been observed that the background 
has less impact on the detector sensitivity to the weak mixing angle.
 It has been bound that with increasing the 
systematic effect($\sigma_{f}$) 10 $\%$ from 5$\%$, the sensitivity of the detector to the weak mixing angle is further reduced to
8.59 $\%$ from 7.33 $\%$ for S/B= 1.0, considering the detector placed at 10 m from the Dhruva reactor core.
%%%%%%%%%%%%%%%%%%%%%%%%%%%
\begin{figure*}[h]
\includegraphics[height=0.3\textwidth]{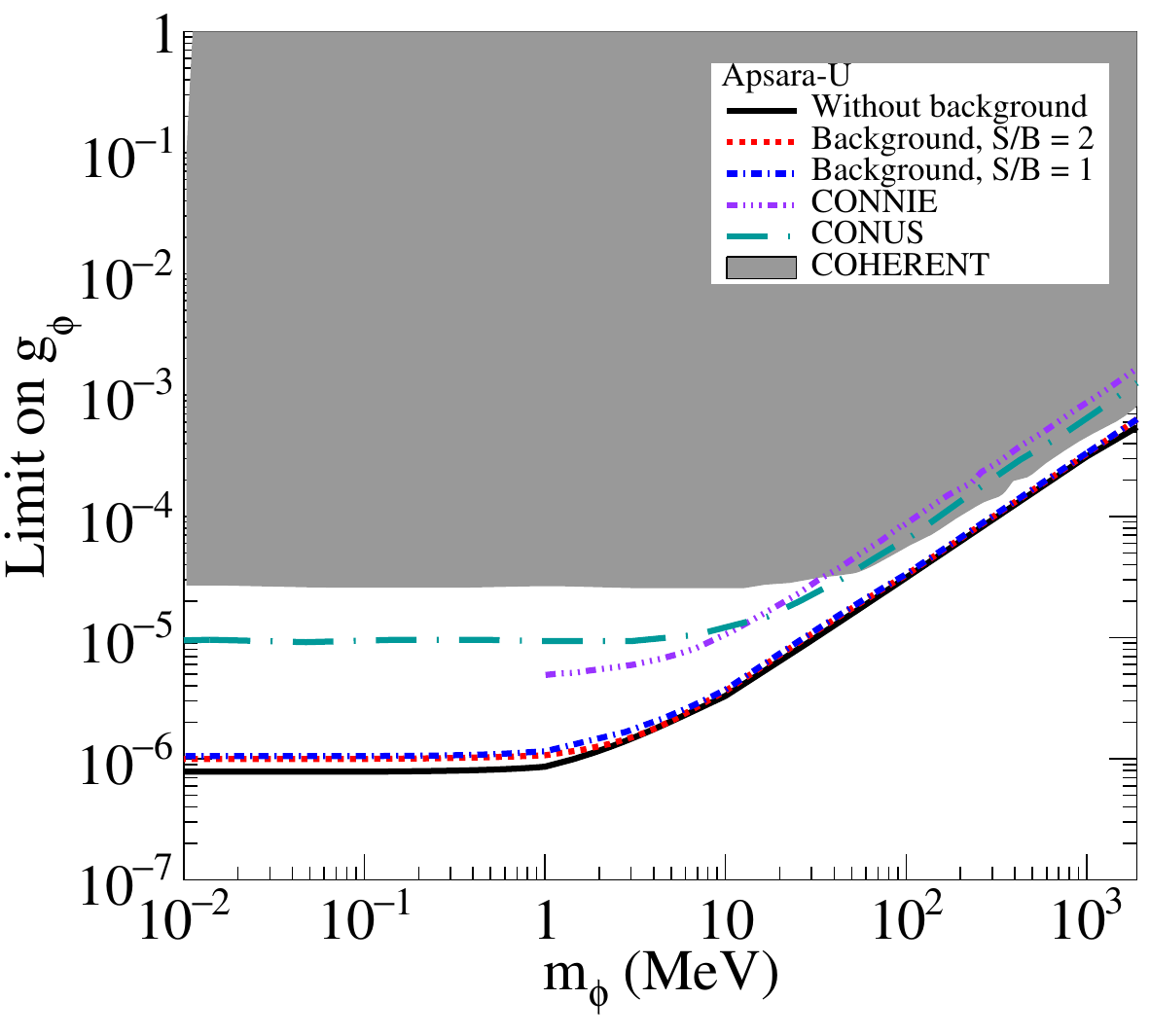}
\includegraphics[height=0.3\textwidth]{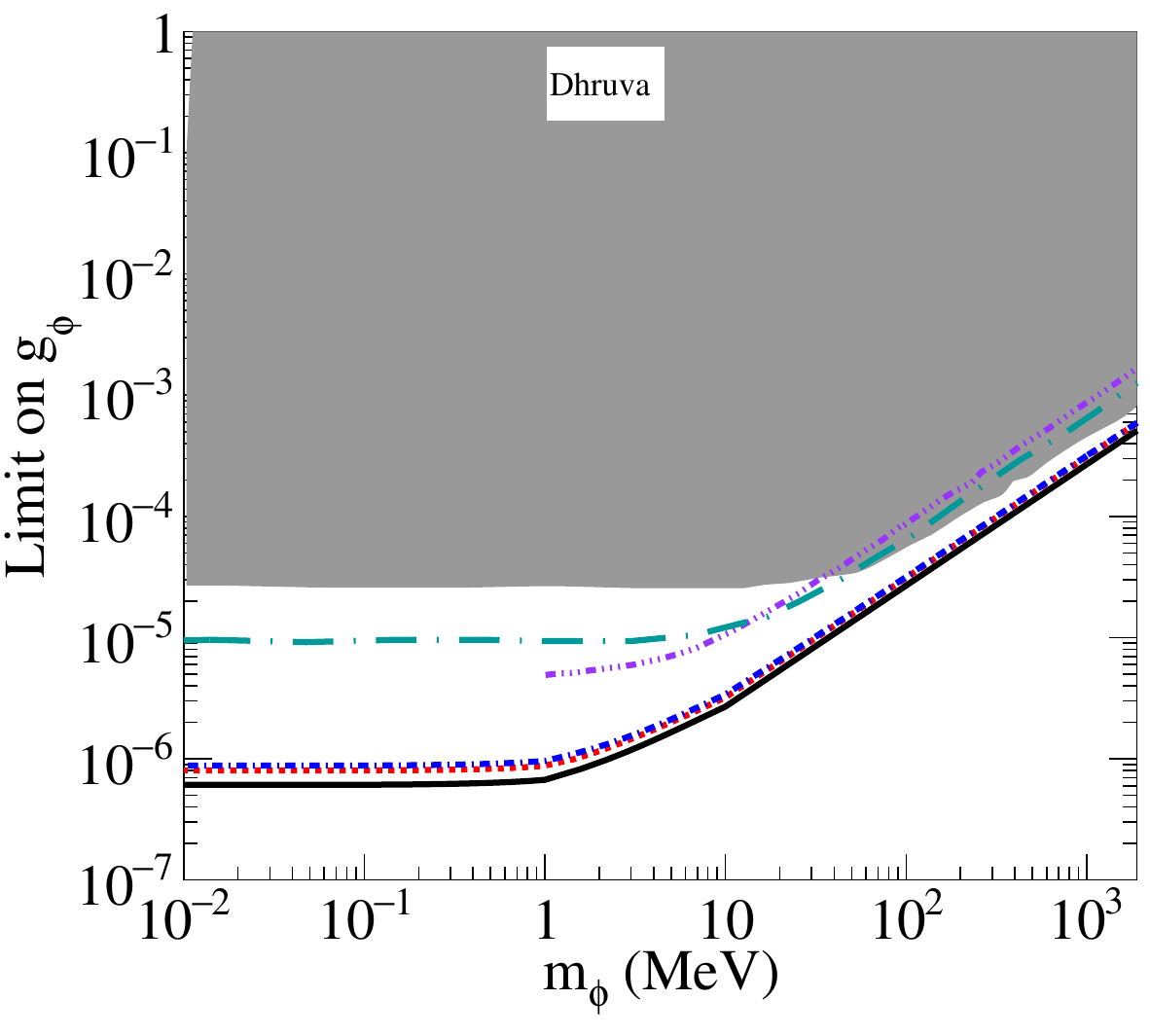}
\includegraphics[height=0.3\textwidth]{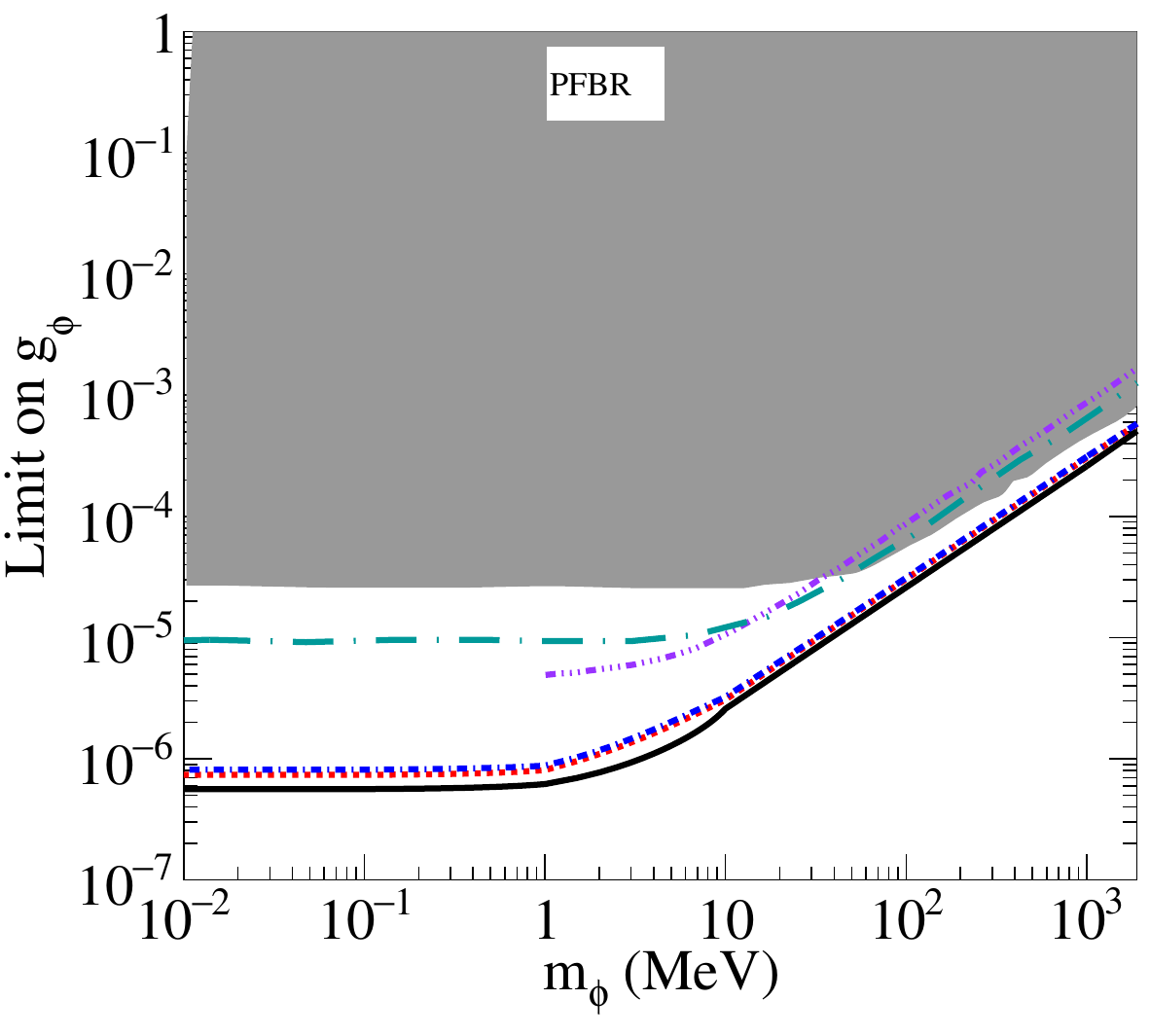}
\includegraphics[height=0.3\textwidth]{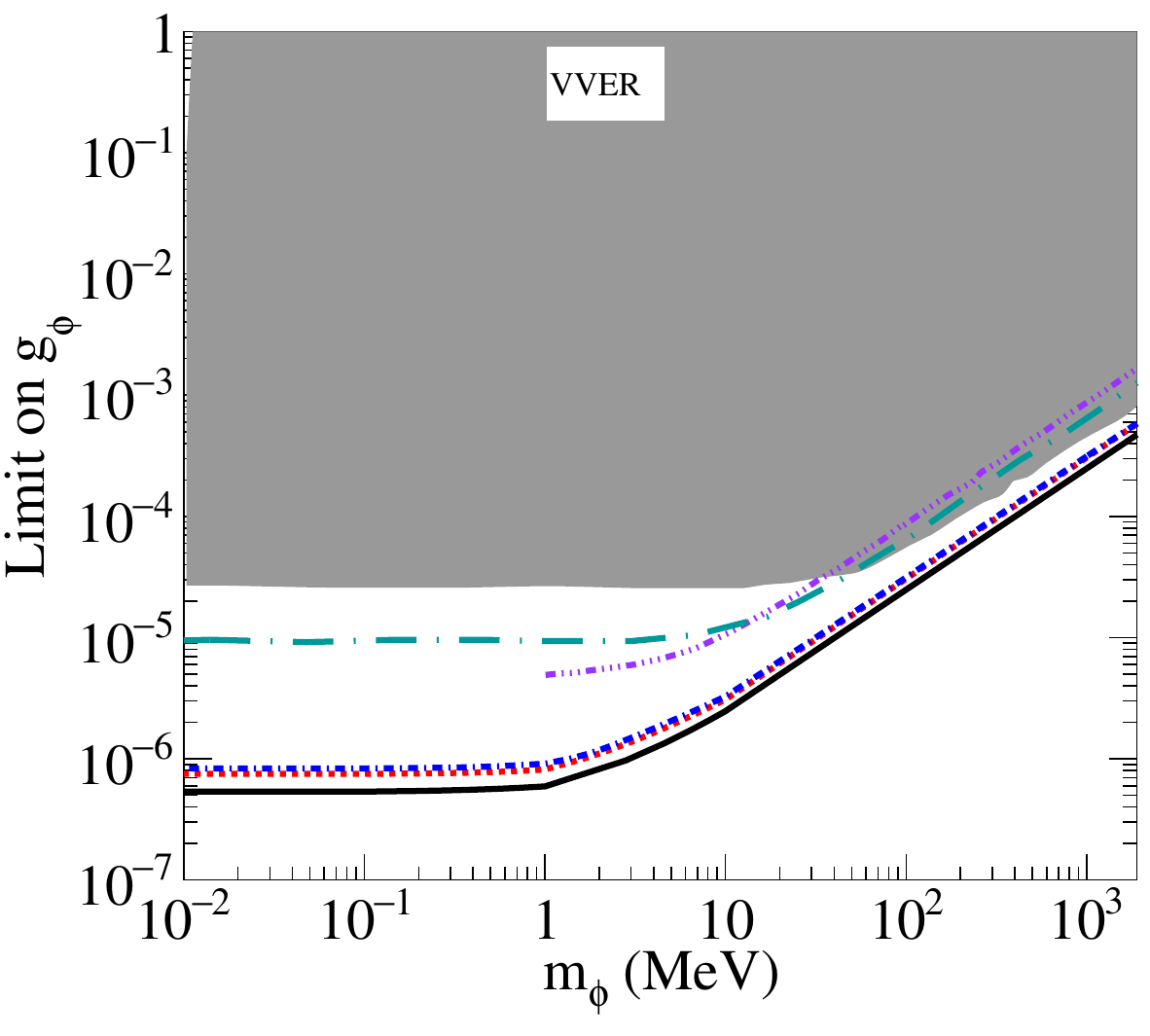}
\vspace{-0.2cm}
\caption{ \label{fig:scalarMediator}The detector sensitivity to the scalar mediator at 90% CL The shaded portion is the excluded region from the COHERENT
group. Results from the CONNIE~\cite{CONNIE:2019xid} and the CONUS~\cite{CONUS:2021dwh}
experiments have been shown for the comparisons.} 
\vspace{-0.5cm}
\end{figure*}
%%%%%%%%%%%%%%%%%%%%%% Fig.1 %%%%%%%%%%%%%%%%%%%%%%%%%%%%%%%%%
%%%%%%%%%%%%%%%%%%%%%%%%%%%
\begin{figure*}[h]
\includegraphics[trim= 0 0 0 9,clip,width=0.34\textwidth]{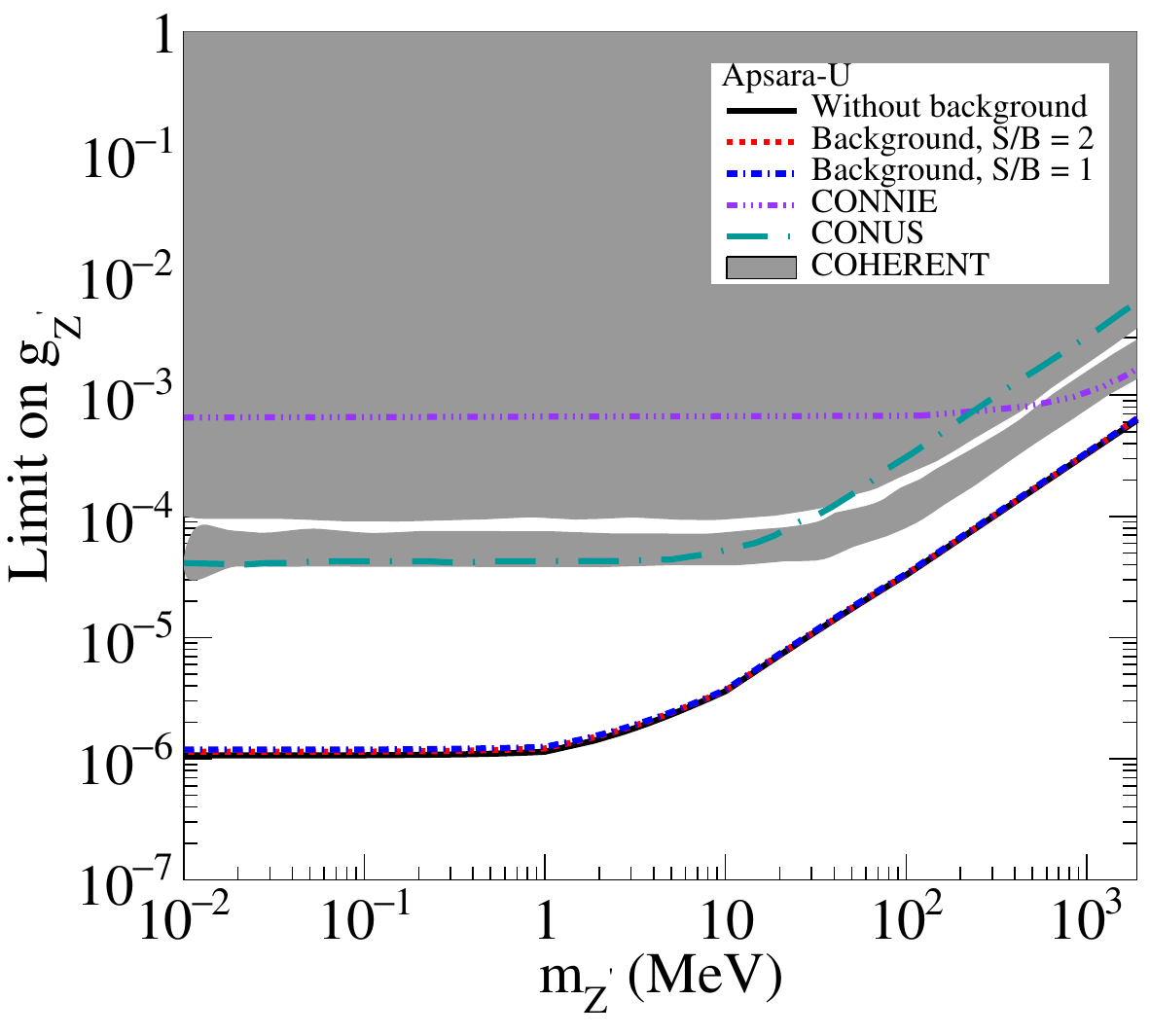}
\includegraphics[trim= 0 0 0 9,clip,width=0.34\textwidth]{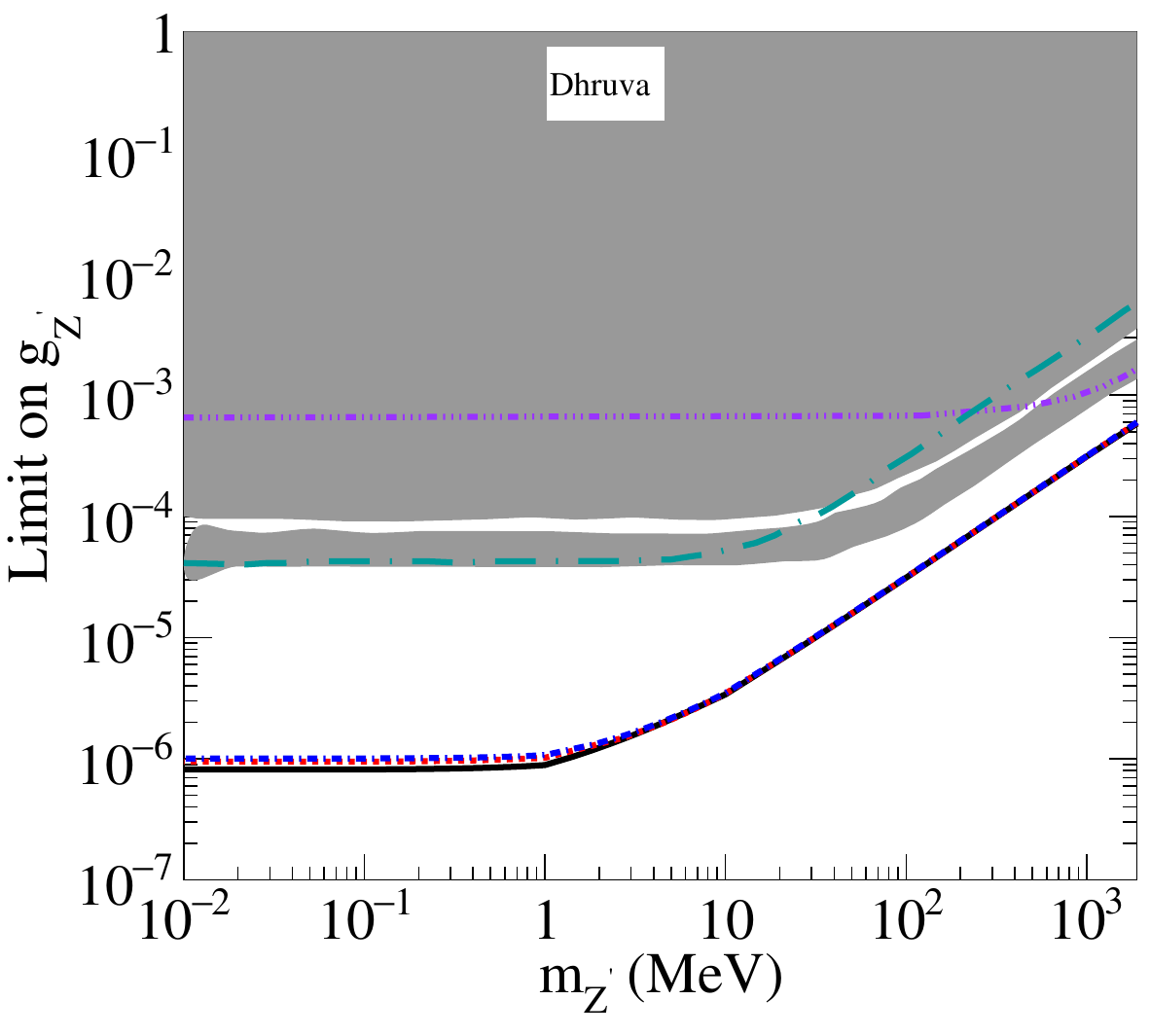}
\includegraphics[trim= 0 0 0 9,clip,width=0.34\textwidth]{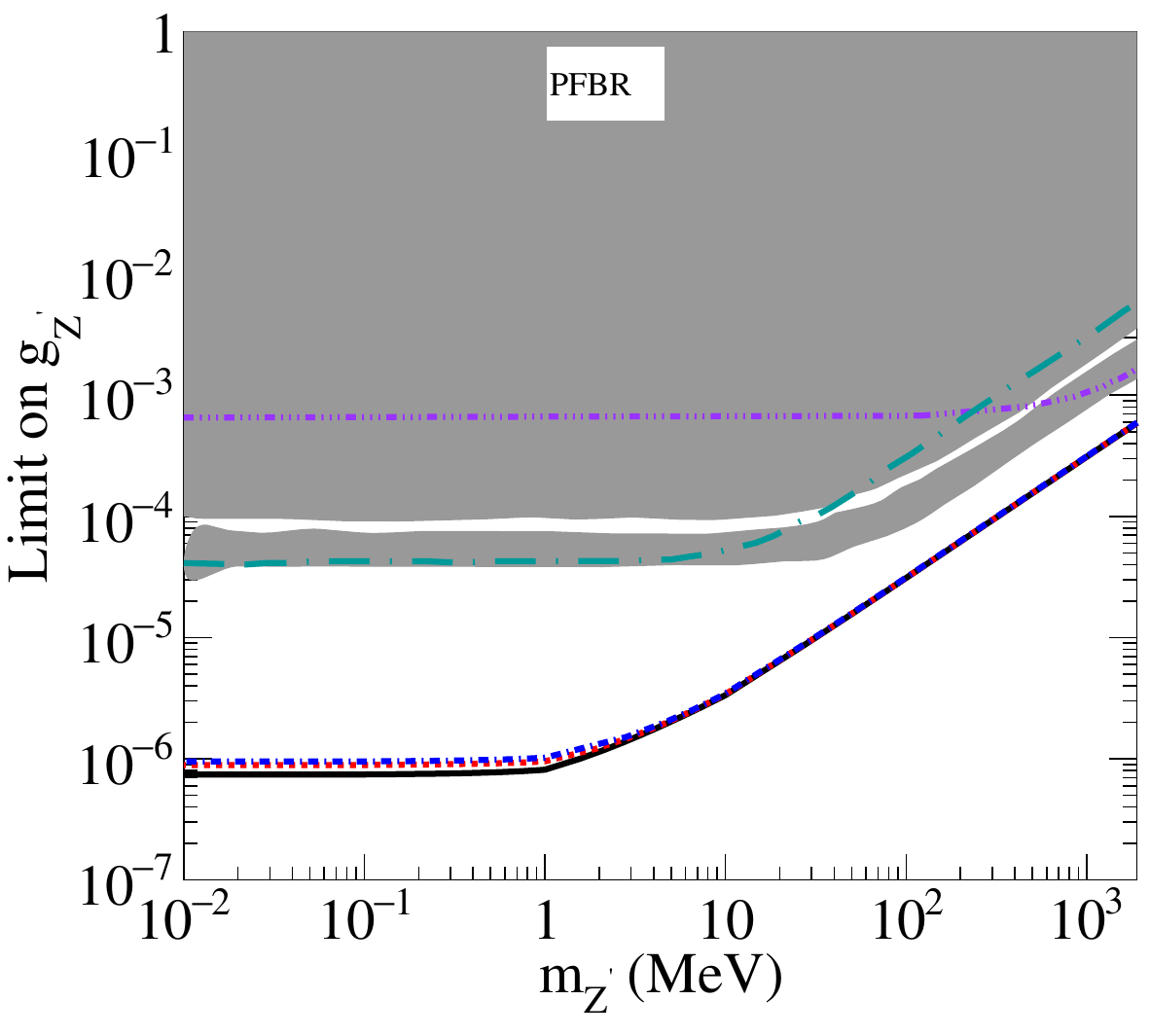}
\includegraphics[trim= 0 0 0 9,clip,width=0.34\textwidth]{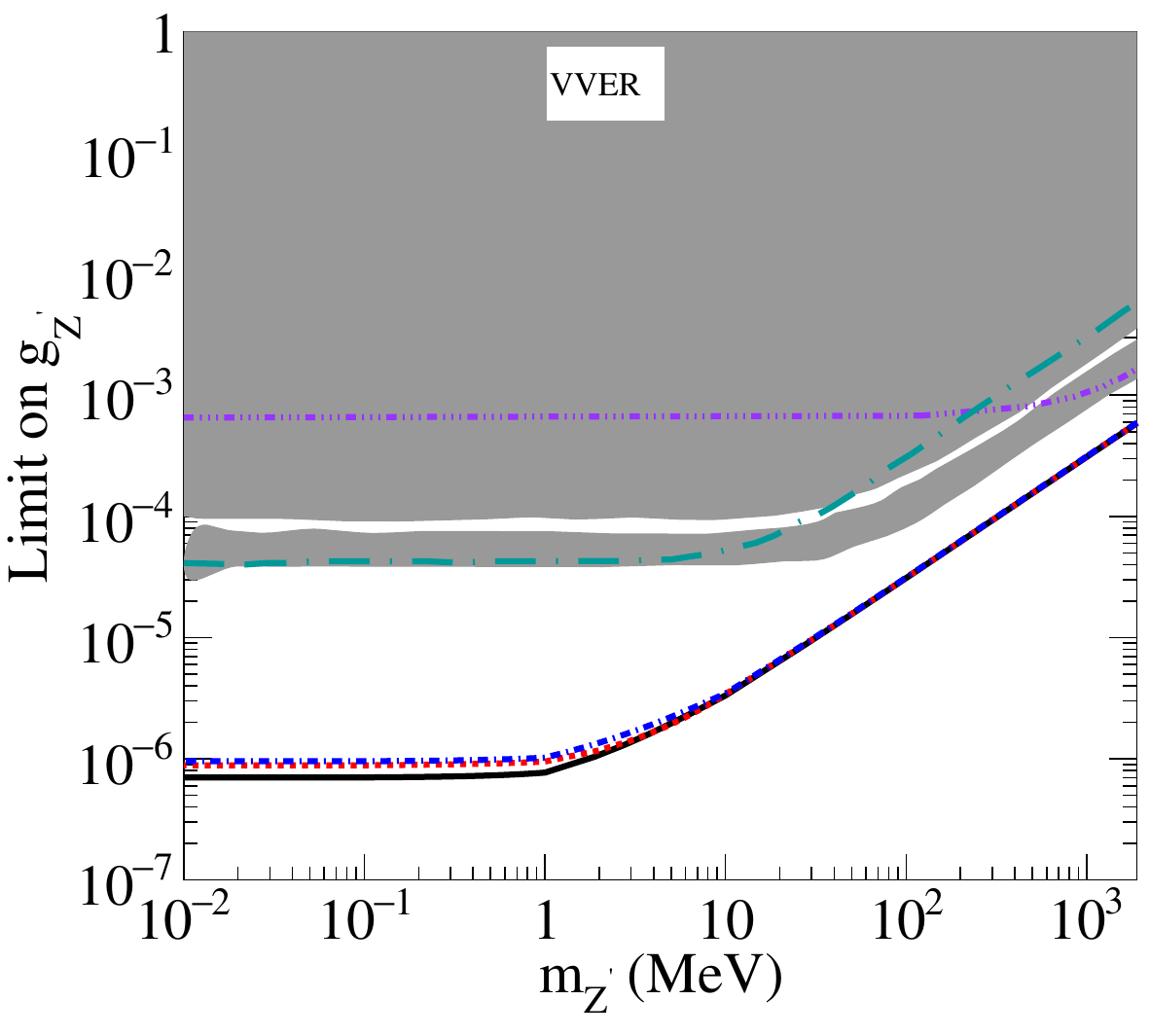}
\caption{ \label{fig:vectorMediator} The detector sensitivity to the vector mediator at 90$\%$C.L. The shaded
portion is the excluded region from the COHERENT group.} 
\end{figure*}
%%%%%%%%%%%%%%%%%%%%%%%%%%%%%%%%%%%%%%%%%%
%%%%%%%%%%%%%%%%%%%%%%
%%%%%%%%%%%%%%%%%%%%
\subsection{Sensitive to the mass of mediators}
%%%%%%%%%%%%%%%%%%%%
The detector's sensitivity to the mediator mass has been extracted by estimating the chi-square 
between the number of events calculated using the SM CE$\nu$NS cross section and with the  deviation from 
SM CE$\nu$NS cross section. Events are estimated with consideration of the SM CE$\nu$NS cross section 
 considered as theoretically predicted ones whereas events are estimated using the cross 
 section influenced by coupling of a new scalar or a vector mediator to fermions are
  considered as simulated measured ones. In $g_{\phi}- m_{\phi}$ plane, Fig.~\ref{fig:scalarMediator} shows the 
  expected sensitivity of the ICNSE detector on the coupling of a new scalar mediator to 
  fermion considering various types of reactor core to the detector distances as 
  well as thermal power. It is assumed that scalar coupling 
    to all SM fermions is universal. 
     At lighter mass region, the detector's sensitivity is independent of mediator mass
     as the interaction cross section depends on only $g_{\phi}$. For heavier scalar
     mass the interaction rate is proportional to $g_{\phi}/m_{\phi}$. 
      At low mass ($\le$10.0 MeV) region, the sensitivity has 
    been reduced due to the presence of background. It is found that detectors have similar sensitivity
     at higher scalar mass region. The ICNSE detectors can exclude most
      of the parameter space as excluded by the COHERENT group.       
       
       An expected potential sensitivity of 
      the ICNSE detector to the mass of the vector mediator is shown in Fig.~\ref{fig:vectorMediator}.
 It is found that the ICNSE detector can exclude most of the parameter space as 
 well as the region excluded by the COHERENT group in $g_{Z^{\prime}}- m_{Z^{\prime}}$ plane.
 At low vector mediator mass region, a similar behavior has been observed as observed for scalar mediators.
   In the presence of background, the detector's sensitivity can deteriorates more at lower mass regions as 
  compared to high mass regions. Results from the CONNIE~\cite{CONNIE:2019xid} and the CONUS~\cite{CONUS:2021dwh} experiments 
      have been shown for the comparisons. 
      At S/B=1 and mediators of mass less than 10 MeV, it has been bound that with increasing the 
systematic effect($\sigma_{f}$) 10 $\%$ from 5$\%$, the sensitivity of detector 
has been reduced about 12$\%$ and  5$\%$
due to the nonstandard interaction caused by scalar and vector mediators,
 respectively with considering the detector placed at a distance of 
10 m from the Dhruva reactor core.
 %%%%%%%%%%%%%%%%%%%%%%%%%%%%%%%%%%%%%%%%%%%%%%%%%%%%%%%%%%%%%%%%
\section{SUMMARY}
\label{sec:summary}
%%%%%%%%%%%%%%%%%%%%%%%%%%%%%
The measurement of low energy recoil nuclei due to the CE$\nu$NS process is 
very challenging. However, the recent measurement of the CE$\nu$NS process by
 the COHERENT group opened a window to further explore physics beyond the 
 standard model of particle physics. Currently, several studies are ongoing, 
 and some studies are proposed to measure many important properties of neutrinos
  using different types of detectors,  taking into account different sources of
   neutrinos.   In this article, we have explored the physics potential of the 
   a sapphire detector in the context of the proposed CE$\nu$NS  scattering experiment
    in India. The study has been performed for a time horizon
     of 1 year employing $\overline{\nu}_e$s produced from 
 the research reactor, which can be further employed for the measurement at the power 
 reactor. It is found that the detector has the potential to measure various BSM physics parameters 
 such  as the neutrino magnetic moment and weak mixing angle using the
CE$\nu$NS process. The detector can limit most of the parameter space 
in  $g_{\phi}- m_{\phi}$  as well as $g_{Z^{\prime}}- m_{Z^{\prime}}$ plane
because of its lower energy thresholds.
  At low mass region of mediators, the background has more effect compared to 
  higher mass regions.  Further, the detector sensitivity  can be improved 
  by reducing background with a proper shielding.
 %%%%%%%%%%%%%%%%%%%%%%%%%%%%%%%%%%%%%%%%%%%%%%%%%%%%%%%%%%%
\section*{ACKNOWLEDGMENTS}
The author thanks V. Jha, D. K. Mishra, Kirtikesh Kumar and, other group members of the ICNSE, BARC for useful suggestions and discussions.
The author also thanks
Dimitrios Papoulias and Luis Flores for giving critical comments on the manuscript. 
%%%%%%%%%%%%%%%%%%%%%%%%%%%%%%%%%%%%%%%%%%%%%%%%%%%%%%%%%%%%%%%%%%%%%%%%%%%%%%%%%%%%%%%%%%%%%%%%%%%%%%%%%%%%%
\bibliography{draft}
\bibliographystyle{apsrev4-1}
\end{document}